\documentclass[11pt]{article}
\usepackage[letterpaper, hmargin=1 in, vmargin=1 in]{geometry}
\usepackage{graphicx}
\usepackage{amsmath, amssymb, amsthm, mathtools}
\usepackage{url}
\usepackage{color}
\usepackage{cite}
\usepackage{xparse}
\usepackage{enumitem}
\usepackage{thmtools}
\usepackage{thm-restate}
\usepackage{hyperref}
\hypersetup{colorlinks=true, linkcolor=[rgb]{0,0,0}, citecolor=[rgb]{0,0,0}, urlcolor=[rgb]{0,0,0}}
\theoremstyle{plain}
\newtheorem{theorem}{Theorem}
\newtheorem{lemma}{Lemma}[section]

\newcommand{\floor}[1]{\left \lfloor #1\right \rfloor}
\newcommand{\ceil}[1]{\left \lceil #1\right \rceil}

\newcommand{\MM}{\mathbb{M}}
\newcommand{\eps}{\varepsilon}
\newcommand{\ST}{\,:\,}

\newcommand{\SP}{\kern+1 pt}

\newcommand{\bIN}{b_{\SP\mathrm{in}}}
\newcommand{\rIN}{r_{\mathrm{in}}}

\DeclareMathOperator{\diam}{diam}

\DeclareMathOperator{\rad}{rad}

\DeclareMathOperator{\rep}{rep}
\DeclareMathOperator{\parent}{par}

\DeclareMathOperator{\Rel}{Rel}

\newcommand{\UU}{\mathcal{U}} 
\newcommand{\EE}{\mathcal{E}} 
\newcommand{\LL}{\mathcal{L}} 
\newcommand{\PP}{\mathcal{P}} 
\newcommand{\NN}{\mathcal{N}} 
\newcommand{\BB}{\mathcal{B}} 
\newcommand{\VV}{\mathcal{V}} 
\newcommand{\DIM}{\mathrm{dim}} 
\newcommand{\ccov}{c_{\mathrm{cov}}} 
\newcommand{\cpack}{c_{\mathrm{pac}}} 
\newcommand{\crel}{c_{\mathrm{rel}}} 
\newcommand{\cgrp}{c_{\mathrm{grp}}} 
\newcommand{\cnet}{c_{\mathrm{net}}} 
\newcommand{\csk}{c^-_{\mathrm{sk}}} 
\newcommand{\Csk}{c^+_{\mathrm{sk}}} 
\newcommand{\cloc}{c_{\mathrm{loc}}} 
\newcommand{\cnear}{c_{\mathrm{nn}}} 
\newcommand{\csep}{c_{\mathrm{sep}}} 
\newcommand{\Ccand}{c_{\mathrm{can}}} 
\newcommand{\Ccoarse}{c_{\mathrm{crs}}} 
\newcommand{\centry}{c_{\mathrm{ent}}} 

\begin{document}

\title{Space--Time Tradeoffs for Proximity Searching in Doubling Spaces%
\thanks{A preliminary version of this work appeared at ESA~2008 \cite{AMV08}.}}

\author{Sunil Arya\thanks{%
        Department of Computer Science and Engineering,
        The Hong Kong University of Science and Technology,
        Clear Water Bay, Kowloon, Hong Kong.
        Email: \texttt{arya@cse.ust.hk}.}
\and	David M. Mount\thanks{%
        Department of Computer Science and Institute 
        for Advanced Computer Studies, University of Maryland, College Park, 
        Maryland 20742.
        Email: \texttt{mount@umd.edu}.}
\and    Antoine Vigneron\thanks{%
        Independent researcher (on leave from INRAE).
        Email: \texttt{antoine.vigneron@polytechnique.org}.}
\and    Jian Xia\thanks{%
        School of Computing and Intelligent Technology,
        Qingdao City University, Qingdao, China.
        Email: \texttt{jian.xia@qdc.edu.cn}.}}


\date{~}

\maketitle

\begin{abstract}
We study space--time tradeoffs for $(1 + \eps)$-approximate nearest-neighbor searching among $n$ sites in an ambient metric space of doubling dimension $\DIM$. Such tradeoffs are known in  Euclidean space, but in the standard black-box model of doubling spaces, the polynomial dependence on $1/\eps$ in query time cannot generally be transferred to storage. To achieve our results, we extend the model to permit stored auxiliary ambient-space points. For $0 < \eps \leq 1/2$ and $2 \leq \gamma \leq 1/\eps$, we prove that there exists a data structure, called an approximate Voronoi diagram (AVD), that uses $2^{O(\DIM)} n \gamma^{\DIM} \log(1/\eps)$ storage and answers queries in $2^{O(\DIM)} \bigl(\log(n\gamma) + (1/(\eps\gamma))^{\DIM}\bigr)$ time. As $\gamma$ increases toward $1/\eps$, the polynomial dependence on $1/\eps$ shifts from query time to storage. A variant of the high-space construction stores one site per AVD cell and achieves $2^{O(\DIM)} n (1/\eps)^{\DIM} \log(1/\eps)$ storage and $2^{O(\DIM)} \log(n/\eps)$ query time. 

The main ingredient is a hierarchical, searchable data structure called a region-DAG, whose cells are metric balls and differences of concentric metric balls. Given access to a \emph{doubling oracle} that covers a given ball in $2^{O(\DIM)}$ time with $2^{O(\DIM)}$ balls of half its radius centered at ambient-space points, both constructions admit randomized preprocessing in expected time $2^{O(\DIM)} n (1/\eps)^{\DIM}$ up to logarithmic factors. Queries are deterministic after preprocessing and do not require the oracle. 

\end{abstract}


\section{Introduction} \label{sec:intro}

Nearest-neighbor searching is a fundamental problem in computational geometry and similarity search, with applications in pattern recognition, information retrieval, machine learning, and other areas. Metric spaces provide a general framework for this problem. A \emph{metric space} is a pair $(\MM, d)$, where $\MM$ is a set and $d$ is a distance function satisfying the properties of a metric (symmetry, positivity, and the triangle inequality). Given a set $S$ of $n$ sites in a metric space $(\MM,d)$, the objective is to preprocess $S$ so that, for any query point $q \in \MM$, we can quickly report a nearest site in $S$ to $q$. Because exact nearest-neighbor searching can be expensive, it is common to allow approximation. For $\eps > 0$, a site $p \in S$ is a \emph{$(1 + \eps)$-approximate nearest neighbor} (ANN) of $q$, if 
\[
    d(q,p) ~ \leq ~ (1 + \eps) d(q,S), 
        \quad \text{where $d(q,S) ~ = ~ \min_{x\in S} d(q,x)$}.
\]

In fixed-dimensional Euclidean space, approximate Voronoi diagrams (AVDs), introduced by Har-Peled \cite{Har01}, provide a geometric approach to this problem. An \emph{$\eps$-AVD} is a searchable data structure that covers the query space with simple regions, called \emph{cells}. It associates each cell with a small set of sites, called \emph{representatives}, such that for every query point in a cell, at least one representative is a $(1 + \eps)$-ANN. A query therefore reduces to locating any cell containing the query point and reporting its closest representative. AVDs can be designed to support storage--query-time tradeoffs, where query times can be reduced at the expense of additional storage \cite{ArM02, AMM02, Har01, AMM09a}.

A natural question is whether this geometric approach extends beyond Euclidean space. For general metrics, low-dimensional structure is commonly expressed through the concept of doubling dimension. After Gupta, Krauthgamer, and Lee~\cite{GKL03} introduced it to this algorithmic setting, doubling dimension became a standard intrinsic measure of metric complexity. 
The \emph{doubling dimension} of a metric space is the smallest value $\DIM$ such that every ball can be covered by at most $2^{\DIM}$ balls of half its radius. Throughout, $\DIM$ denotes the doubling dimension of the ambient space $\MM$, rather than that of the finite site set alone. 

Much of the foundational work on approximate nearest-neighbor searching in doubling spaces uses the \emph{black-box model}~\cite{KrL04, KrL05}. In this model, preprocessing accesses the metric through distance evaluations between sites, and queries access it through distance evaluations between the query point and the sites. Each distance evaluation has unit cost. This model does not assume that points are represented by coordinates, and it provides no mechanism to generate additional points in the ambient space. Hierarchical nets and related structures \cite{HaM06, CoG06} yield data structures with $n2^{O(\DIM)}$ storage and query time 
\[
    2^{O(\DIM)} \log n + (1/\eps)^{O(\DIM)}.
\]
These bounds are independent of the \emph{spread} of the sites, defined as the ratio of the largest to the smallest distance between distinct sites. Cole and Gottlieb's structure also supports insertions and deletions~\cite{CoG06}. 

Unfortunately, the black-box model limits the storage--query-time tradeoffs that can be achieved. Even with unrestricted preprocessing space, doubling metrics exist for which every black-box ANN algorithm requires $(1/\eps)^{\Omega(\DIM)}$ distance evaluations on some query \cite{HaM06}. Consequently, the polynomial dependence on $1/\eps$ cannot be transferred from query time to storage in this model. A separate lower bound shows that, for fixed $\eps$, some doubling metrics require $2^{\Omega(\DIM)} \log n$ distance evaluations \cite{KrL04}. 

These lower bounds leave open what tradeoffs are possible when auxiliary points of the ambient space may be used. This distinction is central to our setting. We assume a relaxed model, the \emph{weakly explicit model}, which permits storing auxiliary ambient-space points and evaluating distances involving them. This model suffices to establish the existence of our data structures; for their efficient construction, we additionally assume access to a \emph{doubling oracle}. Given a metric ball $b$ of radius $r$, the oracle returns a collection of $2^{O(\DIM)}$ balls of radius $r/2$ whose union covers $b$. Their centers are points of the ambient metric space and need not belong to either $b$ or $S$. We assume that the oracle runs in $2^{O(\DIM)}$ time, and distances involving the center points generated by the oracle can be computed in constant time.

In this paper, we establish a parameterized storage--query-time tradeoff for AVDs in general doubling metric spaces in the weakly explicit model. We also show that these AVDs can be constructed efficiently given a doubling oracle. Our approach provides an explicit hierarchical cover with efficient point location, whose cells are metric balls or differences of two concentric metric balls. Each terminal region stores a small list of representative sites such that, for every query point in the region, at least one stored site is a $(1 + \eps)$-ANN. 

Har-Peled and Kumar~\cite{HaK13} considered AVDs in metric spaces earlier, studying ANN queries restricted to a subspace of low doubling dimension. If we take this subspace to be the entire ambient space, their projection operation becomes the identity. Under a comparable ball-covering assumption, their static construction achieves $n(1/\eps)^{O(\DIM)}$ storage, $2^{O(\DIM)}\log(n/\eps)$ query time, and $n(1/\eps)^{O(\DIM)}\log n$ expected preprocessing time. This construction searches auxiliary points using a net-tree, rather than explicitly constructing and searching a cover by simple geometric cells. They also give a separate online AVD construction in a weaker-access model. Its regions are balls and differences of balls, not necessarily concentric, with one representative valid throughout each region. However, efficient point location in these regions remains unresolved. Neither construction establishes the parameterized storage--query-time tradeoff considered here.

Our main theorem, stated below, presents a continuous tradeoff parameterized by $\gamma$. For $2 \leq \gamma \leq 1/\eps$, increasing $\gamma$ trades additional storage for a smaller polynomial term in the query-time bound. It also gives a variant of the high-space construction that stores exactly one representative per cell. We assume that all points and numerical quantities used by the data structure can be stored in constant space, and that distance evaluations and basic numerical operations take constant time. (See Section~\ref{sec:preliminaries} for further details.) 

\begin{restatable}{theorem}{avdTheorem} \label{thm:avd}
Let $(\MM,d)$ be a metric space of doubling dimension $\DIM$, let $S \subseteq \MM$ be a set of $n \geq 2$ sites, and let $0 < \eps \leq 1/2$. In the weakly explicit model, the following hold.
\begin{enumerate}[label=$(\roman*)$, ref=\roman*]
\item \label{item:avd-tradeoff} For every $2 \leq \gamma \leq 1/\eps$, there exists an $\eps$-approximate Voronoi diagram for $S$ with at most $2^{O(\DIM)} n \gamma^{\DIM} \log(1/\eps)$ cells and at most $2^{O(\DIM)} (1/(\eps\gamma))^{\DIM}$ representatives per cell, achieving
\[
    \text{Storage:~}
        2^{O(\DIM)} n \gamma^{\DIM} \log(1/\eps)
    \qquad\text{and}\qquad
    \text{Query time:~}
        2^{O(\DIM)} \left( \log(n\gamma) + \left(\frac{1}{\eps\gamma}\right)^{\!\DIM} \right). 
\]

\item \label{item:avd-single-rep} There exists an $\eps$-approximate Voronoi diagram for $S$ with at most $2^{O(\DIM)} n (1/\eps)^{\DIM} \log(1/\eps)$ cells and one representative per cell, using the same asymptotic storage and answering queries in $2^{O(\DIM)} \log(n/\eps)$ time.
\end{enumerate}
Given access to a doubling oracle, both constructions admit Las Vegas randomized preprocessing in expected time 
\[ 
    2^{O(\DIM)} n (1/\eps)^{\DIM} \log(1/\eps)\log(n/\eps). 
\] 
\end{restatable}

The two extremes illustrate the tradeoff in Theorem~\ref{thm:avd}(\ref{item:avd-tradeoff}). For $\gamma = 2$, the structure uses $n 2^{O(\DIM)} \log(1/\eps)$ storage and has query time $2^{O(\DIM)} \bigl( \log n + (1/\eps)^{\DIM} \bigr)$. For fixed doubling dimension, the storage is only $O(n \log(1/\eps))$. At the opposite extreme, setting $\gamma = 1/\eps$ gives storage $2^{O(\DIM)} n (1/\eps)^{\DIM} \log(1/\eps)$ and query time $2^{O(\DIM)} \log(n/\eps)$. Thus, the polynomial dependence on $1/\eps$ shifts from query time to storage. For fixed doubling dimension, the query time is logarithmic in $n/\eps$, a bound that cannot generally be achieved uniformly over $\eps$ in the black-box model. Ignoring logarithmic factors, the product of these storage and query-time bounds is $2^{O(\DIM)} n (1/\eps)^{\DIM}$ for every $\gamma$ in the tradeoff range.

Theorem~\ref{thm:avd}(\ref{item:avd-single-rep}) gives a variant of the high-space construction that stores exactly one representative per cell, closer in spirit to Har-Peled's original definition of the AVD~\cite{Har01}. This variant matches the high-space endpoint's asymptotic storage, query-time, and expected preprocessing bounds. 

Our main technical contribution is a hierarchical, searchable cover called a \emph{region-DAG}. Euclidean AVD constructions rely on hierarchical box decompositions based on quadtrees and their variants, but an abstract doubling metric space supplies neither coordinates nor a canonical analog of their disjoint recursive decomposition. The region-DAG replaces the box-decomposition tree with a hierarchical, possibly overlapping cover adapted to the metric geometry. Its cells are metric balls or differences of concentric metric balls. The leaf cells satisfy separation conditions that permit small representative sets, and the hierarchy supports efficient point location despite overlaps.

Other related work considers complementary metric settings and access models. Abraham, Chechik, Krauthgamer, and Wieder developed compact data structures for bounded-degree planar graph metrics, which need not have bounded doubling dimension~\cite{AbCKW15}. Near-neighbor-preserving dimension reduction has been obtained for doubling subsets of $\ell_1$ \cite{EMP23}, and Filtser introduced labeled nearest-neighbor searching using locality-sensitive orderings, with results that include doubling spaces~\cite{Fil23}. More recently, Lu and Tao studied proximity-graph search in doubling metrics, obtaining graph-size and query bounds that depend on the spread of the data~\cite{LuT25}. These works address complementary access models, ambient spaces, or variants of the nearest-neighbor problem. Our focus is an explicitly represented geometric AVD with efficient point location and a parameterized storage--query-time tradeoff in general bounded-doubling ambient metric spaces.

\subsection{Overview of Techniques} \label{sec:overview}

Our approach generalizes Euclidean AVD constructions, which are based on quadtree-like decompositions into cells. In order to allow for a small number of nearest-neighbor representatives, the decomposition is refined so that each leaf cell satisfies a separation condition with respect to the input set. A query locates a leaf cell containing the query point and returns the representative nearest to that point. The decomposition may be constructed using a box-decomposition tree \cite{AMNSW94} or a compressed quadtree \cite{Har11a}. The fatness of the underlying boxes and the explicit spatial hierarchy are central to the analysis.

To extend this approach to doubling metrics, we replace the Euclidean partition with a cover based on metric balls, using auxiliary ambient-space points permitted by the weakly explicit model. A net-tree~\cite{HaM06} provides a compressed hierarchy of site clusters that guides the construction. Its covering and packing properties determine the scales and locations of site-centered balls, while a doubling oracle supplies additional balls to cover the intervening space. Applying the packing bound in the ambient metric gives an exponent of exactly $\DIM$ on the ratio of scales. Shrinking operations bypass large gaps between scales by introducing doughnut cells, avoiding dependence on the spread of the site set. The resulting hierarchy is represented by a directed acyclic graph, called the \emph{region-DAG}, whose cells may overlap and whose nodes may have multiple parents.

The leaves of the region-DAG satisfy one of two separation conditions, which are controlled by two parameters $\gamma$ and $\beta$, where $\gamma, \beta \geq 2$. For a simple leaf whose cell is a metric ball $b$, all input sites other than a site possibly at its center lie outside $\gamma b$, the concentric ball whose radius is $\gamma$ times that of $b$. A doughnut leaf, the difference of two concentric metric balls, satisfies both inner and outer separation conditions: nearby input sites lie deep inside its inner ball, with separation controlled by a parameter $\beta$, while all remaining sites lie outside the $\gamma$-expansion of its outer ball. These separation conditions underlie the construction of representative sets valid throughout each leaf cell. Thus, although leaf cells may overlap, any leaf cell containing the query point can be used to answer the query. We fix $\beta  = \Theta(1/\eps)$ and use $\gamma$ to control the storage--query-time tradeoff.

For leaves that contain a site, this site itself serves as the representative. For a doughnut leaf, the common center of its two defining balls serves as the representative, even though it lies outside the cell. The challenge is to handle representatives for empty simple leaves. A metric range sketch supplies a separated set of \emph{candidate sites} that approximates all sites relevant to queries in the leaf. Using the doubling oracle, we construct a net for the leaf whose points serve as auxiliary \emph{helpers}, and select a nearest candidate for each helper. The selected sites form the leaf's representative list. A query in the leaf scans this list, so its number of entries determines the representative-search portion of the query time. 

To bound total storage, we adapt the WSPD-based charging argument of~\cite{AMM09a}. A well-separated pair decomposition (WSPD) of the sites is a collection of pairs of well-separated subsets such that every two distinct sites lie on opposite sides of exactly one pair. An empty simple leaf with $k$ candidates charges $k-1$ distinct WSPD pairs, each of which separates two of its candidates. A WSPD pair may be charged by several leaves, but candidate separation and locality restrict the radius scales of those leaves. At each such scale, the region-DAG's local-multiplicity property bounds the number of leaves that can charge the pair. The charge count, together with the bound on the number of leaves, bounds the total number of candidates and hence the total representative storage. The charging scheme is used only in the analysis and need not be computed.

For the single-representative variant, we apply the high-space construction with a smaller approximation parameter. For each empty simple leaf, we retain only the representative nearest to its center. The leaf's separation from the sites ensures that this representative remains a $(1 + \eps)$-ANN throughout the leaf. The other terminal regions already have one representative each. 

Section~\ref{sec:preliminaries} introduces the metric tools used in the construction. Section~\ref{sec:region-dag} defines the region-DAG and states its properties, and Section~\ref{sec:region-dag-construction} presents its construction, point-location algorithm, and complexity bounds. Section~\ref{sec:avd} uses the region-DAG to obtain the AVD space--time tradeoff and the single-representative variant. Section~\ref{sec:conc} concludes with open problems. The appendices give implementation details for the net-tree tools and the construction of hierarchical ball covers.

\section{Preliminaries}\label{sec:preliminaries} 

Throughout the paper, $(\MM,d)$ denotes a metric space in the weakly explicit model, and $S \subseteq \MM$ is a set of $n \geq 2$ \emph{sites}. The space $\MM$ may be discrete or continuous, finite or infinite, bounded or unbounded. For $x \in \MM$ and $r \geq 0$, let
\[
    B(x, r) ~ := ~ \{y \in \MM : d(x, y) \leq r \}
\]
denote the closed \emph{metric ball} of radius $r$ centered at $x$. Throughout, a ball is specified by a center--radius pair, and different pairs are treated as distinct even if they define the same subset of $\MM$. For a ball $b = B(x,r)$, write $\rad(b) = r$ and, for $\eta > 0$, let $\eta b = B(x,\eta r)$. For $X \subseteq \MM$, define $d(q,X) := \inf_{x \in X} d(q,x)$, with $d(q,\emptyset) = +\infty$. Throughout, we use $\log$ to denote the base-2 logarithm.

The \emph{doubling dimension} of $\MM$, denoted by $\DIM(\MM)$, is the smallest value such that every ball of positive radius can be covered by at most $2^{\DIM(\MM)}$ balls of half its radius. We set $\DIM(\MM) = +\infty$ if no such finite value exists. We say that $\MM$ is a \emph{doubling space} if $\DIM(\MM)$ is finite. Throughout, we assume that $\MM$ is doubling and write $\DIM = \DIM(\MM)$. Since $\MM$ contains at least two distinct sites, we have $\DIM \geq 1$. We treat $\DIM$ as an asymptotic quantity, and in expressions such as $2^{O(\DIM)}$, the constants hidden in the exponents are absolute. Unless stated otherwise, all other hidden constants are independent of $n$, $\eps$, $\beta$, $\gamma$, and $\DIM$. In complexity bounds, logarithmic factors of the input size are understood to be at least one. 

The following packing bound is a straightforward consequence of the doubling condition. It bounds the number of separated points that can lie in a metric ball. 

\begin{lemma}[Doubling packing bound]\label{lem:doubling-packing}
Let $r$ and $\delta$ be positive scalars with $\delta \leq r$, and let $P \subseteq B(x, r)$ be a set in a doubling space of dimension $\DIM$ such that $d(p, p') \geq \delta$ for all distinct $p,p'\in P$. Then $|P| \leq (4 r/\delta)^{\DIM}$.
\end{lemma} 

\begin{proof} 
Set $k := \floor{\log_2 (2 r/\delta)} + 1$. Iterating the defining doubling property $k$ times gives a cover of $B(x,r)$ by at most $2^{k \cdot \DIM}$ balls of radius $r/2^k$. Each covering ball has diameter at most $2 r/2^k < \delta$, so it contains at most one point of $P$. Therefore, 
\[ 
    |P| 
        ~ \leq ~ 2^{k \cdot \DIM} 
        ~ \leq ~ (4 r/\delta)^{\DIM}.
\] 
\end{proof} 

In the \emph{weakly explicit model}, the data structure may store arbitrary points of $\MM$, including auxiliary centers outside $S$. Each stored point or number occupies constant storage. Distance evaluations between sites, stored auxiliary points, and query points take constant time. We also assume constant-time arithmetic, comparisons, logarithms, fixed-base exponentiation, and floor operations on represented numbers. 

The model permits auxiliary points to be stored but does not itself provide a procedure for generating them; this is handled by a function called a doubling oracle. Given a ball $b = B(z, r)$, a \emph{doubling oracle} returns a collection of $2^{O(\DIM)}$ balls of radius $r/2$ whose union covers $b$. We assume the oracle runs in $2^{O(\DIM)}$ time. Access to this oracle is an additional assumption used for efficient construction, not for queries. Note that this oracle is not an intrinsic element of the weakly explicit model, and in our results, we point out explicitly where access to a doubling oracle is assumed.

Nets are a fundamental concept in metric spaces. Given a subset $X \subseteq \MM$ and $\rho > 0$, a set $\NN \subseteq \MM$ is called a \emph{$\rho$-net for $X$} if:
\begin{itemize} \setlength{\itemsep}{-0.5 ex}\setlength{\parsep}{0 pt}
\item every point of $X$ lies within distance $\rho$ of some point of $\NN$, and 

\item every two distinct points of $\NN$ are at distance at least $\cnet\rho$, where $\cnet > 0$ is a fixed absolute constant chosen sufficiently small.
\end{itemize}

We do not require $\NN \subseteq X$. Net points may be auxiliary ambient-space centers. Every subset of a metric space admits such a net (see, for example, \cite{BBI01, Hei01, KrL04}). Lemma~\ref{lem:hierarch-ball-cover} below proves the existence of small nets and covering hierarchies, and it shows how to construct them assuming access to a doubling oracle.

\subsection{Net-Trees} \label{sec:net-trees}

In real $d$-dimensional space, structures such as grids and quadtrees can be used to access data efficiently. The corresponding structures in metric spaces are nets and net-trees. Given a set $P$ of sites in a metric space $\MM$, a net-tree for $P$ is a rooted, compressed tree structure whose leaves are associated with individual sites of $P$, and whose internal nodes represent clusters of sites based on metric balls. As with quadtrees, scales increase geometrically toward the root. There are numerous implementations of the net-tree in the literature; we use the construction of Har-Peled and Mendel~\cite{HaM06} and summarize its relevant properties here.

A net-tree storing a set of sites $P$ in doubling space $\MM$ has the following properties. Leaves correspond one-to-one with the sites. For each node $u$, let $P_u$ denote the sites associated with the leaves of its subtree.  If $u$ is not the root, let $\parent(u)$ denote its parent. Each node stores a \emph{representative} site, $\rep(u) \in P_u$, and is associated with a \emph{level}, denoted $\lambda(u)$. Internal nodes have integer levels, and leaves have level $-\infty$. Levels strictly increase toward the root, but edges may skip arbitrarily many levels. Every internal node has at least two and at most $2^{O(\DIM)}$ children, and its representative is inherited from one of its children. It follows that a net-tree storing $m$ sites has $O(m)$ nodes.

\begin{figure}[htbp]
  \centering\includegraphics[scale=0.40]{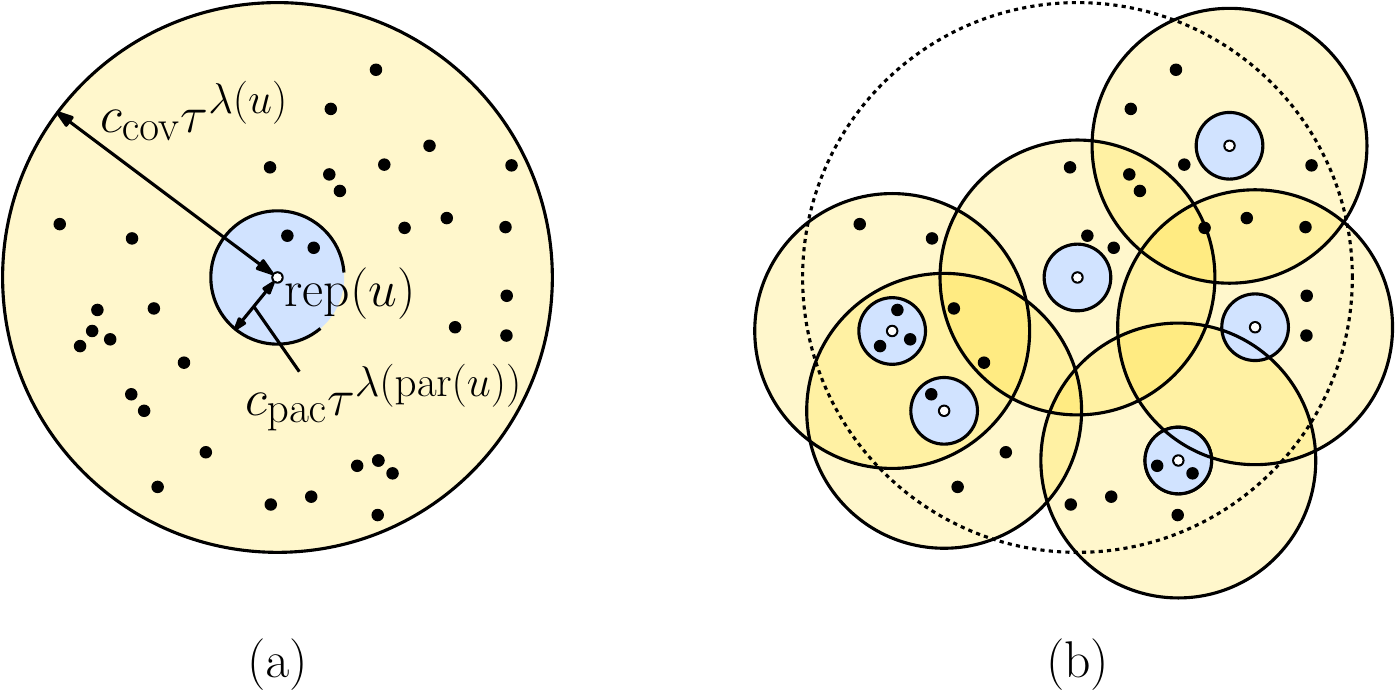}
  \caption{Cover and packing properties of the net-tree for (a) a node $u$ and (b) its children.} \label{f:net-tree}
\end{figure}

Let $\tau > 1$ be the absolute scale base of the construction ($\tau = 11$ in~\cite{HaM06}), and set $\tau^{-\infty} = 0$. The covering and packing properties of Har-Peled and Mendel~\cite[Definition~2.1]{HaM06} give absolute constants $\ccov \geq 1$ and $\cpack > 0$ such that
\begin{align}
    P_u
        & ~ \subseteq ~ B\bigl(\rep(u), \ccov \tau^{\lambda(u)}\bigr), \label{eq:nt-cover} \\
    P_u
        & ~ \supseteq ~ P \cap B\bigl(\rep(u), \cpack\tau^{\lambda(\parent(u))}\bigr),
        \quad\text{for nonroot $u$} \label{eq:nt-pack}
\end{align}
(see Figure~\ref{f:net-tree}). The first property bounds the extent of a subtree cluster; the second ensures that every site sufficiently close to its representative belongs to that cluster.

A useful notion for identifying nodes of a given scale is that of a cut. First, it will be convenient to add a virtual parent of the root of level $+\infty$. For an integer $k$, the \emph{level-$k$ cut} is the set of nodes
\[
    \VV_k
        ~ := ~ \bigl\{ u \ST \lambda(u) < k \leq \lambda(\parent(u)) \bigr\}.
\]
Thus, the cut selects the first node of level below $k$ on each path from the root to a leaf. If the root level is below $k$, the cut consists only of the root. Clearly, the sets $P_u$, for $u \in \VV_k$, partition $P$.

The doubling dimension of $P$ with its induced metric is at most $2\cdot\DIM$. Indeed, an ambient ball of radius $r$ can be covered by at most $2^{2\cdot\DIM}$ balls of radius $r/4$. For each such ball that meets $P$, a ball of radius $r/2$ centered at any point of $P$ in it covers its intersection with $P$. It follows from Har-Peled and Mendel's construction~\cite[Theorem~3.1]{HaM06} that a net-tree on $P$ can be built in expected time $2^{O(\DIM)} m \log m$, where $m = |P|$. The construction uses distance evaluations only, requires no doubling oracle, and is independent of the spread.

A \emph{scale-location query}, specified by a site $x \in P$ and an integer $k$, returns the unique node in $\VV_k$ whose subtree contains the leaf corresponding to $x$. Its implementation is given in Appendix~\ref{app:net-tree-tools}.

\begin{restatable}{lemma}{scaleLocation} \label{lem:scale-location}
Let $T$ be a net-tree on $m$ sites. Then $T$ can be augmented in $O(m)$ time using $O(m)$ additional storage to answer scale-location queries in $O(\log m)$ time.
\end{restatable}

\subsection{Finite-Metric Tools} \label{sec:finite-metric-tools}

Our algorithms will employ three further operations on finite inputs in $\MM$. Like scale location, they use only distance evaluations, require no doubling oracle, and generate no new ambient-space points. All bounds are independent of the spread. The approximate nearest-neighbor result is due to Har-Peled and Mendel~\cite[Theorem~4.4]{HaM06}. The fixed-scale grouping and metric range-sketching interfaces are proved in Appendix~\ref{app:net-tree-tools} using their net-tree machinery.

\begin{lemma}[Approximate nearest-neighbor searching] \label{lem:net-tree-ann}
Let $P \subseteq \MM$ be a nonempty set of $m$ sites. In $2^{O(\DIM)} m \log m$ expected time, one can construct a data structure that uses $2^{O(\DIM)} m$ storage and, given a query point $q$ and a parameter $0 < \delta \leq 1$, returns a $(1 + \delta)$-ANN of $q$ in worst-case time $2^{O(\DIM)} \log m + (1/\delta)^{O(\DIM)}$.
\end{lemma}

As a useful special case, setting $\delta = 1$ yields a factor-$2$ approximation in $2^{O(\DIM)} \log m$ time. Our next lemma derives fixed-scale grouping from the net-tree construction of Har-Peled and Mendel~\cite{HaM06}, with repeated input locations handled explicitly in the appendix. Fixed-scale grouping accepts an indexed family of points, with repetitions allowed, and assigns every input occurrence to a selected center. 

\begin{restatable}[Fixed-scale grouping]{lemma}{fixedScaleGrouping} \label{lem:fixed-scale-grouping} 
Let $(p_i)_{i=1}^m$ be a nonempty indexed family of points in $\MM$, with repetitions allowed, and let $s > 0$. One can construct a set of center indices $A \subseteq \{1,\dots,m\}$ and an assignment $\pi \colon \{1,\dots,m\} \to A$ such that:
\begin{enumerate} \setlength{\itemsep}{-0.5 ex}\setlength{\parsep}{0 pt} 
\item[$(i)$] $d(p_i, p_{\pi(i)}) \leq s$ for every $i$, and $\pi(i) = i$ for every $i \in A$, and
\item[$(ii)$] for distinct $i,j \in A$, $d(p_i,p_j) \geq \cgrp s$, where $\cgrp > 0$ is an absolute constant. 
\end{enumerate} 
The expected construction time is $2^{O(\DIM)} m \log m$. 
\end{restatable} 

The final operation returns a separated set of sites approximating the sites in a query ball at a prescribed resolution. Returned sites may lie in a constant-factor enlargement of the ball. 

\begin{restatable}[Metric range sketching]{lemma}{metricRangeSketch} \label{lem:metric-range-sketch}
Let $P \subseteq \MM$ be a nonempty set of $m$ sites. In $2^{O(\DIM)} m \log m$ expected time, one can construct a data structure that occupies $2^{O(\DIM)} m$ storage and supports the following operation. Given $q \in \MM$, $r > 0$, $0 < \eta \leq r$, and a site $p_0 \in P$ satisfying $d(q,p_0) \leq 2 r$, return a set $C \subseteq P$ such that:
\begin{itemize} \setlength{\itemsep}{-0.5 ex}\setlength{\parsep}{0 pt}
\item \emph{(Coverage)} for every $p \in P \cap B(q,r)$, there is an $x \in C$ satisfying $d(p,x) \leq \eta$,
\item \emph{(Separation)} distinct sites in $C$ are separated by at least $\csk \eta$, and
\item \emph{(Locality)} $C \subseteq B(q,\Csk r)$.
\end{itemize}
Here $\csk > 0$ and $\Csk \geq 1$ are absolute constants. The worst-case query time is
\[
    O(\log m) + 2^{O(\DIM)} + O(|C|).
\]
\end{restatable}

\subsection{Hierarchical Ball Covers}\label{sec:hierarchical-ball-covers}

In this section, we present an important element of our region-DAG construction. We combine the doubling oracle with fixed-scale grouping to construct small nets and bounded-degree covering hierarchies. Grouping is applied to an entire level, so centers at that level are separated even when they originate from different parent balls. We choose the constant in the net definition so that $0 < \cnet \leq \min\{1,\cgrp\}/4$.

\begin{restatable}[Nets and hierarchical ball covers]{lemma}{hierarchBallCover} \label{lem:hierarch-ball-cover}
Let $b = B(z,r) \subseteq \MM$, where $r > 0$. For any $\rho$, $0 < \rho \leq r$, there exists a $\rho$-net $\NN$ for $b$ and a ball-covering DAG rooted at $b$ with the following properties.
\begin{enumerate}
\item[$(i)$] $\NN \subseteq B(z, r + \rho)$ and $|\NN| ~ \leq ~ 2^{O(\DIM)} (r/\rho)^{\DIM}$.

\item[$(ii)$] Every internal ball is covered by at most $2^{O(\DIM)}$ children of half its radius. All node centers lie in $B(z,3 r)$. All leaves have the same radius in $(\rho/2,\rho]$. 

\item[$(iii)$] The DAG has at most $2^{O(\DIM)} (r/\rho)^{\DIM}$ nodes and edges.
\end{enumerate}
Given a doubling oracle, the net and DAG can both be constructed in expected time
\[
    2^{O(\DIM)} (r/\rho)^{\DIM} \log(2 r/\rho).
\]
\end{restatable}

The construction applies the doubling oracle twice to each current ball, then uses fixed-scale grouping on all candidate centers at the next level. Grouping across the entire level gives separated centers, while assignments preserve covering by children and bounded outdegree. A packing argument bounds the number of level-$j$ centers by $2^{O(\DIM)} 2^{j \cdot \DIM}$, and summing over the levels gives the stated bound on the number of nodes. The full proof is given in Appendix~\ref{app:hierarchical-ball-covers}.

\section{The Region-DAG} \label{sec:region-dag}

In this section, we introduce the \emph{region-DAG}, a central data structure in our AVD construction. For a set $S$ of $n$ sites in a metric space $(\MM,d)$ of finite doubling dimension $\DIM$, it provides a space-efficient, searchable hierarchical cover of a designated root ball. In their Euclidean AVD construction, Arya, Malamatos, and Mount~\cite{AMM09a} use a box-decomposition tree whose leaf cells satisfy separation conditions with respect to the sites. The region-DAG plays an analogous role in our construction, but its cells may overlap. To construct it, we use the net-tree of Har-Peled and Mendel~\cite{HaM06}. The net-tree organizes the sites into clusters with packing and covering guarantees, but its leaves represent individual sites rather than regions of the query space. We use this cluster hierarchy to build a cover whose leaf cells satisfy the separation conditions needed for our AVD.

Each node represents a region, called a \emph{cell}, defined using metric balls. The children of an internal node cover its cell, but a child's cell need not be contained in its parent's cell. Unlike the net-tree, the balls defining the cells need not be centered at points of $S$ and are generally obtained by applying the doubling oracle. The construction takes two independent parameters $\gamma \geq 2$ and $\beta \geq 2$, which govern different separation properties of the cells and, in turn, affect storage and query time. The parameter $\gamma$ determines separation from sites outside an enlarged cell ball, while $\beta$ determines separation from sites deep inside a doughnut cell's inner ball. In the AVD construction, we set $\beta = \Theta(1/\eps)$ and vary $\gamma$ to obtain the storage--query-time tradeoff.

We designate a node of the region-DAG as its \emph{root}. The associated cell, called the \emph{root ball}, is a ball centered at a point of $S$ that contains $S$ (and indeed significantly more). Given a node $u$ whose associated cell is a ball $b$, invoking the doubling oracle results in a set of $2^{O(\DIM)}$ balls of half the radius that cover $b$. We can use these balls to define $u$'s children. Intuitively, we could iterate this process until the cells are so small and so well separated from the points of $S$ that we can efficiently infer the approximate nearest neighbor of any query point in the cell from a small set of representative points of $S$. Unfortunately, a straightforward application of this process would yield storage and depth bounds that depend on the spread of the point set. Instead, we use the net-tree to identify isolated site clusters and introduce doughnut cells to bypass the intervening gaps between scales. 

The region-DAG has two node types, depending on the cell's form. A \emph{simple cell} is a single metric ball $b$, and a \emph{doughnut cell} is the set difference of two concentric balls, $b_O \setminus b_I$, called the \emph{outer} and \emph{inner} balls, respectively. The nodes associated with these cells are called \emph{simple nodes} and \emph{doughnut nodes}, respectively. The inner ball's radius is at most half the outer ball's radius and may be much smaller. 

The \emph{size} of a simple cell is the radius of its defining ball, and the \emph{size} of a doughnut cell is the radius of its outer ball. A node's size is the size of its associated cell. A cell is \emph{empty} if it contains no point of $S$, and \emph{nonempty} otherwise. If the region-DAG contains an edge from a node $u$ to a node $v$, then $v$ is a \emph{child} of $u$. A child's cell need not be contained in its parent's cell. A node with no children is a \emph{leaf}; otherwise, it is \emph{internal}. 

\subsection{Region-DAG Properties} \label{sec:region-dag-props}

Before describing its construction, let us present the salient properties of the region-DAG. 

\begin{enumerate}[label=(\roman*), ref=\roman*]
\item \label{prop:root-ball} \textbf{Root ball.} There is a simple node whose associated ball $b_0$ is centered at a point of $S$ and satisfies 
\[
    S ~ \subseteq ~ \frac{1}{\beta} \, b_0.
\]
We designate one such node as the \emph{root}, and call $b_0$ the \emph{root ball}. The root ball defines the domain of point-location queries, which may start at other stored nodes. The root need not be a source node of the DAG.

\item \label{prop:sep-leaf} \textbf{Leaves.} Leaves are simple or doughnut nodes and satisfy the following separation properties. 

\begin{enumerate}[label=(\alph*), ref=\alph*]
\item \label{prop:sep-simple} \textbf{Simple leaf.} Let $b$ be the associated ball. Then $\gamma b$ either contains no point of $S$ (see Figure~\ref{f:leaves}(a)) or contains exactly one point of $S$ at the center of $b$ (see Figure~\ref{f:leaves}(b)). There is an absolute constant $\cloc \geq 1$, independent of the dimension and the separation parameters, such that whenever $S \cap \gamma b = \emptyset$, 
\begin{equation}
    S \cap \cloc \gamma b ~ \neq ~ \emptyset. \label{eq:empty-leaf-locality}
\end{equation}
In this case, the leaf stores a site in $S \cap \cloc \gamma b$. 

\begin{figure}[htbp]
  \centering\includegraphics[scale=0.40]{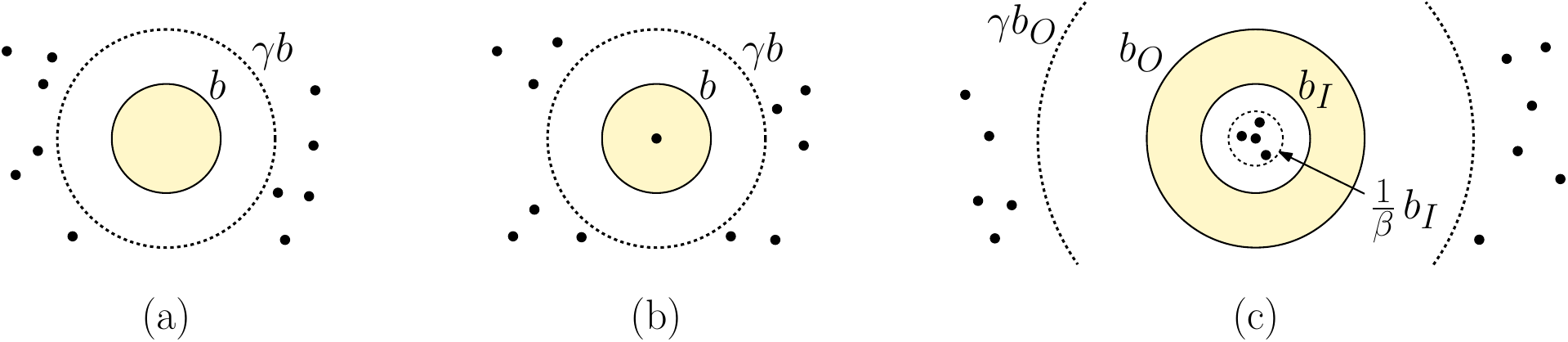}
  \caption{Leaf types: (a) empty simple leaf, (b) nonempty simple leaf, and (c) doughnut leaf.} \label{f:leaves}
\end{figure}

\item \label{prop:sep-doughnut} \textbf{Doughnut leaf.} Let $b_O$ and $b_I$ be the outer and inner balls of the cell associated with a doughnut leaf (see Figure~\ref{f:leaves}(c)). The balls $b_O$ and $b_I$ are concentric, and their common center is a point $p\in S$. Moreover,
\[
    S \cap \gamma b_O
        ~ \subseteq ~ \frac{1}{\beta} \, b_I.
\]
In particular, $p \in \beta^{-1} b_I$, and, since $\beta^{-1} b_I \subseteq b_I$, the doughnut cell $b_O \setminus b_I$ contains no points of $S$. Thus, every point of $S$ lies either deep inside $b_I$ or outside $\gamma b_O$. The parameters $\beta$ and $\gamma$ control separation from the inner and outer sites, respectively.  
\end{enumerate}

\item \label{prop:leaf-multiplicity} \textbf{Local multiplicity of empty simple leaves.}
For every $x \in \MM$, $s>0$, and $\kappa \geq 1$, the number of empty simple leaves whose balls have radii in $[s,2 s)$ and centers in $B(x,\kappa \gamma s)$ is at most 
\[
    2^{O(\DIM)}(\kappa\gamma)^{\DIM}. 
\]
Distinct leaf nodes are counted separately, even if their defining balls coincide. The constant hidden in $O(\DIM)$ is absolute.

\item \label{prop:covering} \textbf{Covering by children.} Every internal node is simple (that is, doughnut nodes arise only as leaves). The associated cell is covered by its children's cells. Internal nodes are of two types: \emph{splitting nodes} and \emph{shrinking nodes}. Let $u$ be an internal node whose associated ball $b$ has radius $r$. 

\begin{enumerate}[label=(\alph*), ref=\alph*]
\item \label{prop:split} If $u$ is a splitting node, then it has $2^{O(\DIM)}$ children. Every child is simple and has size in the interval $[r/32, r/2]$ (see Figure~\ref{f:children}(a)). 

\begin{figure}[htbp]
  \centering\includegraphics[scale=0.40]{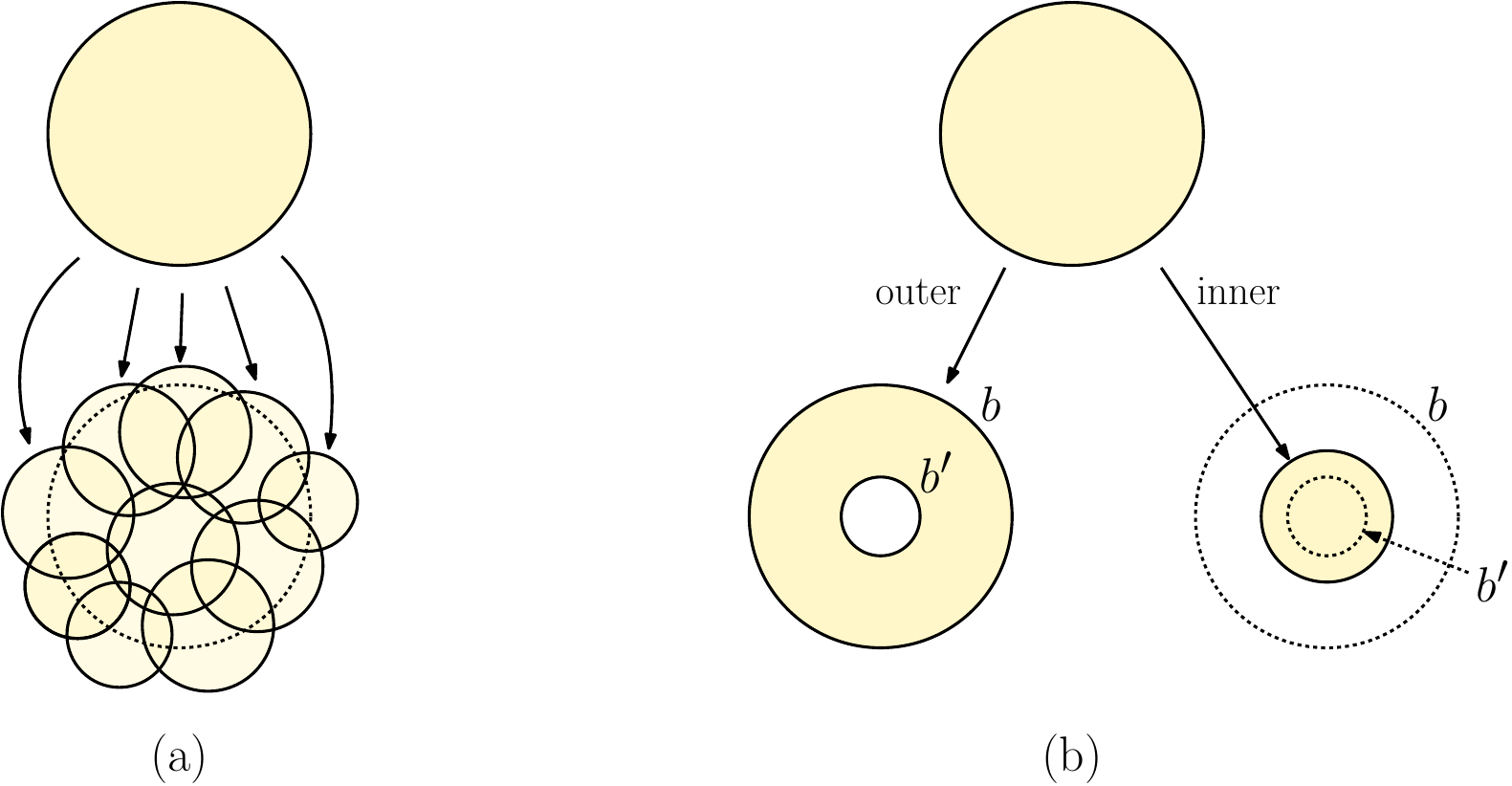}
  \caption{The child cells associated with (a) a splitting node and (b) a shrinking node.} \label{f:children}
\end{figure}

\item \label{prop:shrink} If $u$ is a shrinking node, then it has two children, and all balls defining their cells are concentric with $b$. One child is a doughnut leaf with outer ball $b$ and inner ball $b'$ of radius at most $r/2$. The other child is simple, with ball radius in $[\rad(b'),r/2]$. We call these the \emph{outer child} and \emph{inner child} of $u$, respectively (see Figure~\ref{f:children}(b)). Every doughnut leaf is the outer child of a shrinking node. 
\end{enumerate}

\item \label{prop:pt-loc} \textbf{Scale-sensitive point location.} Given a query point $q$ in the root ball $b_0$ and a scale parameter $\Lambda$ with $0 \leq \Lambda \leq \rad(b_0)$, the following operation can be performed in $2^{O(\DIM)} \log(n\gamma)$ time.
\begin{itemize}
\item If $\Lambda = 0$, return a leaf whose associated cell contains $q$.
\item If $\Lambda > 0$, return one of the following. 
\begin{enumerate}[label=(\alph*), ref=\alph*]
\item \label{prop:pl-leaf} A leaf $u$ of size at least $\Lambda$ whose associated cell contains $q$ (see Figure~\ref{f:point-location-1}(a)). 

\item \label{prop:pl-intern} An internal node $u$ and a simple child $v$, with associated balls $b$ and $b'$, respectively, such that $q \in b \cap b'$ and $\rad(b') < \Lambda \leq \rad(b)$. (See Figure~\ref{f:point-location-1}(b) and~(c) for the cases where $u$ is a splitting and shrinking node, respectively.) 
\end{enumerate}
\end{itemize}

\begin{figure}[htbp]
  \centering\includegraphics[scale=0.40,page=1]{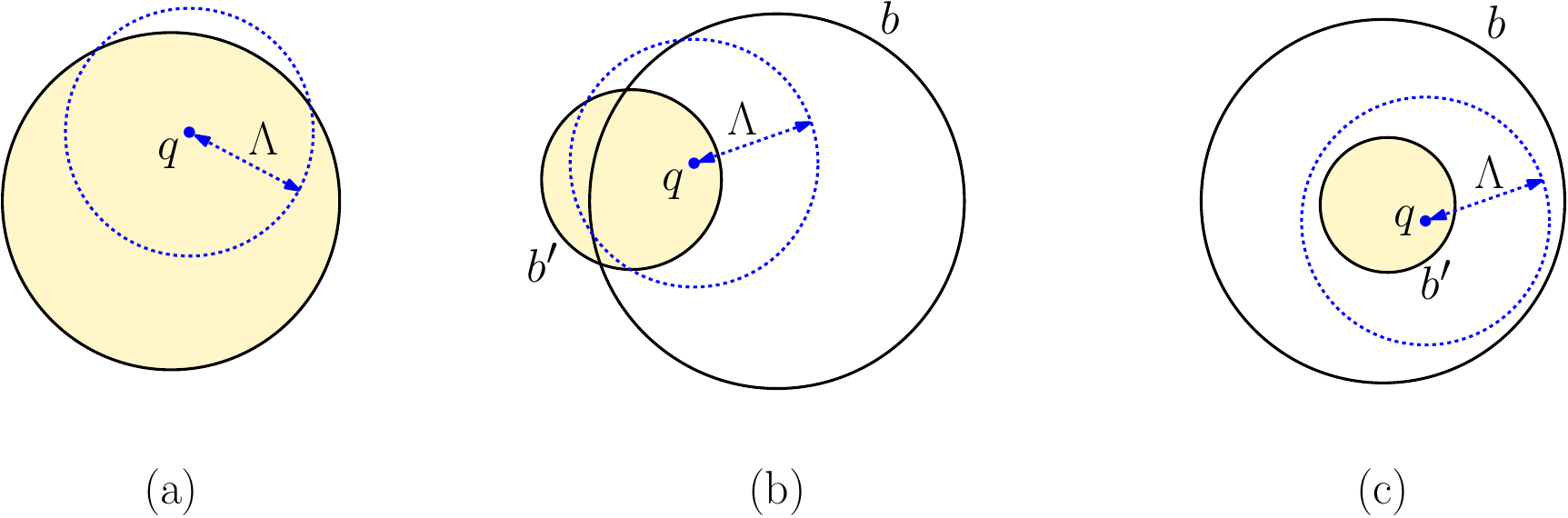}
  \caption{Scale-sensitive point location for $\Lambda > 0$.} \label{f:point-location-1}
\end{figure}
\end{enumerate}

Along every edge leading to a simple node, the associated ball's radius decreases by at least a factor of two. Every edge that does not decrease the outer radius leads to a doughnut leaf. Thus, the graph is acyclic. In a finite region-DAG, property~(\ref{prop:covering}) implies that every node's cell is covered by leaf cells. In particular, the leaf cells cover the root ball. 

The root-ball margin and leaf-separation conditions allow us to assign small representative sets to the exterior and leaf regions. Child covering and point location let us find an applicable leaf for any query in the root ball. A query outside the root ball belongs to the exterior region. These are the structural ingredients of the approximate Voronoi diagram in Section~\ref{sec:avd}. 

\begin{theorem}\label{thm:region-dag}
Given a set $S$ of $n \geq 2$ points in a metric space $\MM$ of doubling dimension $\DIM$ and parameters $\gamma \geq 2$ and $\beta \geq 2$, there exists a region-DAG satisfying properties~$(\ref{prop:root-ball})$--$(\ref{prop:pt-loc})$ with at most 
\[
        2^{O(\DIM)} n \gamma^{\DIM} \log(\beta\gamma)
\]
nodes and using $2^{O(\DIM)} n \gamma^{\DIM} \log(\beta\gamma)$ storage. Given access to a doubling oracle for $\MM$, the region-DAG can be
constructed by a Las Vegas randomized algorithm in expected time
\[
        2^{O(\DIM)} n \gamma^{\DIM} \log(\beta\gamma) \log(n\beta\gamma).
\]
The expectation is over the random choices used in the net-tree-based
preprocessing of $S$ and in the fixed-scale grouping calls used to
construct the hierarchical ball covers.
\end{theorem}

Property~(\ref{prop:covering}) bounds the outdegree by $2^{O(\DIM)}$, so the edge count satisfies the same asymptotic bound as the node count. The auxiliary net-tree and search structures also fit within the stated storage bound.

\section{Region-DAG Construction} \label{sec:region-dag-construction}

We now prove Theorem~\ref{thm:region-dag}. The construction uses a net-tree to identify isolated clusters and metric range sketches to select nearby sites. The nonempty simple cells are defined by selected site-centered balls, called \emph{principal balls}. These are chosen from a family of \emph{canonical balls} defined by net-tree cuts, and their radii are \emph{dyadic}, that is, powers of two. Each principal ball becomes a splitting node, a shrinking node, or a singleton simple leaf. A splitting node uses smaller principal balls to cover nearby sites and empty \emph{derived balls} to cover the remaining space. Refining the derived balls for $O(\log\gamma)$ levels yields leaves whose $\gamma$-expansions contain no sites.

We initially retain all canonical balls classified as splitting. Their radii fall into short bands near the scales of net-tree nodes and their parents, giving $O(n\log(\beta\gamma))$ such balls. Processing them requests at most $2^{O(\DIM)}$ principal children per ball. A requested ball outside the initial family becomes either a singleton leaf or a shrinking node whose simple inner child already belongs to that family. Thus, no further splitting balls are introduced. Identical principal balls share a single region-DAG node. The total number of principal nodes is therefore $2^{O(\DIM)} n \log(\beta\gamma)$. Each splitting node also generates at most $2^{O(\DIM)} \gamma^{\DIM}$ derived nodes, yielding the theorem's bound on the total number of nodes.

For point location, an approximate nearest-neighbor query and the requested scale determine a query scale. A scale-location query then identifies a stored splitting ball. If its radius is much larger than the query scale, a principal child either supplies the answer or determines it through a shrinking step. Otherwise, a descent of $O(\log\gamma)$ steps suffices. 

\subsection{Principal Balls and their Classification} \label{sec:principal-node-classification}

Let $T$ be a net-tree on $S$, with the notation and constants of Section~\ref{sec:net-trees}. For a node $u$ of $T$, let $S_u$ denote the sites in its subtree, and write 
\[
    a^+_u ~:=~ \ccov\tau^{\lambda(u)}
    \qquad\text{and}\qquad
    a^-_u ~:=~ \cpack\tau^{\lambda(\parent(u))},
\]
where $a^+_u = 0$ for leaves and $a^-_u = +\infty$ for the root. The covering and packing properties give
\begin{equation}
    S\cap B(\rep(u),a^-_u) ~\subseteq~ S_u ~\subseteq~ B(\rep(u),a^+_u). \label{eq:principal-cluster-bounds}
\end{equation}
For the root, the first containment is automatic because $S_u=S$.

For $r > 0$, set
\begin{equation}
    k(r) ~ := ~ \floor{\log_\tau\frac{r}{8\ccov}} + 1. \label{eq:principal-cut-level}
\end{equation}
A pair $(u,r)$ is \emph{canonical} if $u \in \VV_{k(r)}$ and $r$ is a power of two. Its associated ball $b(u,r) := B(\rep(u), r)$ is also called \emph{canonical}. Equivalently, for a nonroot node $u$, a dyadic radius $r$ makes $(u,r)$ canonical precisely when 
\begin{equation} 
    8 a^+_u ~\leq~ r ~<~ 8 a^+_{\parent(u)}. \label{eq:principal-canonical-range} 
\end{equation} 
For the root, only the lower bound applies. For a site $x\in S$ and a dyadic radius $r$, let $u(x,r)$ denote the unique node of $\VV_{k(r)}$ whose subtree contains $x$. Since $a^+_{u(x,r)} \leq r/8$, the covering property gives 
\begin{equation}
    d(x, \rep(u(x,r))) ~ \leq ~ \frac{r}{8}. \label{eq:principal-canonicalization}
\end{equation}
These statements include the case in which the cut consists only of the root.

Canonical balls are potential principal balls, but only a selected subset is \emph{retained} in the region-DAG. We classify them to determine the types of their nodes when retained. Consider the test
\begin{equation}
    r ~ \geq ~ 16\beta a^+_u
        \qquad\text{and}\qquad
    r ~ \leq ~ \frac{a^-_u}{\gamma}. \label{eq:principal-shrink-test}
\end{equation}
A canonical ball $b(u,r)$ satisfying this test is classified as \emph{singleton} if $u$ is a leaf and as \emph{shrinking} otherwise. If the test fails, the ball is classified as \emph{splitting}. Intuitively, the test ensures that the ball's radius is significantly larger than the covering radius $a_u^+$ of $u$ and significantly smaller than its packing radius $a_u^-$. When retained, a singleton ball becomes a simple leaf, and a shrinking or splitting ball becomes an internal node of the corresponding type.

For a canonical ball satisfying the test, Eqs.~\eqref{eq:principal-cluster-bounds} and~\eqref{eq:principal-shrink-test} imply that
\begin{equation}
    S \cap \gamma b(u,r)
        ~\subseteq~ S_u
        ~\subseteq~ B(\rep(u),a^+_u). \label{eq:principal-isolated-cluster}
\end{equation}
Thus, all sites in the $\gamma$-expansion belong to a cluster lying deep inside the ball. For the root, the first containment is automatic. 

Initially, we retain all canonical splitting balls. Let $\BB_0$ denote this family. We also retain the canonical balls requested as principal children when these balls are processed. Every retained ball outside $\BB_0$ is singleton or shrinking, and every shrinking node's simple inner child already belongs to $\BB_0$, as shown below. Each retained canonical ball is represented by a single principal node. 

\begin{lemma}\label{lem:initial-principal-balls}
The family $\BB_0$ contains $O(n\log(\beta\gamma))$ balls and can be enumerated in that many arithmetic operations.
\end{lemma}

\begin{proof}
Fix a node $u$ of $T$. If a canonical ball $b(u,r)$ fails the first inequality of Eq.~\eqref{eq:principal-shrink-test}, then by Eq.~\eqref{eq:principal-canonical-range}, we have
\[
    8 a^+_u 
        ~ \leq ~ r 
        ~ < ~ 16\beta a^+_u. 
\]
For internal $u$, this interval contains $O(\log\beta)$ powers of two, and for a leaf, it is empty. If $u$ is not the root, canonicity also implies $k(r)\leq\lambda(\parent(u))$, and hence
\[
    r 
        ~ < ~ 8\ccov\tau^{\lambda(\parent(u))} 
        ~ = ~ \frac{8\ccov}{\cpack} a^-_u. 
\]
Failure of the second inequality therefore restricts $r$ to
\[
    \frac{a^-_u}{\gamma} 
        ~ < ~ r
        ~ < ~ \frac{8\ccov}{\cpack} a^-_u. 
\]
This interval contains $O(\log\gamma)$ powers of two. For the root, the second inequality never fails.

For each node $u$, enumerate the dyadic radii in these two intervals, avoiding duplicates, and retain precisely those pairs $(u,r)$ that are canonical and classified as splitting. Canonicity is tested in constant time by checking $\lambda(u) < k(r) \leq \lambda(\parent(u))$, with the virtual-parent convention for the root (see Section~\ref{sec:net-trees}). Every ball in $\BB_0$ is obtained in this way. The net-tree has $O(n)$ nodes, so the stated bounds follow.
\end{proof}

\subsection{Simple Leaves and Shrinking Nodes} \label{sec:shrinking-nodes}

Consider a retained canonical ball $b=b(u,r)$ satisfying Eq.~\eqref{eq:principal-shrink-test}, and set $x=\rep(u)$. If $u$ is a leaf of $T$, then $S_u = \{x\}$, so Eq.~\eqref{eq:principal-isolated-cluster} gives $S \cap \gamma b=\{x\}$. The corresponding region-DAG node is therefore a valid simple leaf.

Otherwise, define
\[
    b_I ~ := ~ B(x,\beta a^+_u), \qquad
    \rIN ~ := ~ 2^{\ceil{\log_2 (4\beta a^+_u)}}, \qquad\text{and}\qquad
    \bIN ~ := ~ B(x,\rIN).
\]
Create a doughnut leaf with cell $b\setminus b_I$ as the outer child and make the principal node associated with 
$\bIN$ the simple inner child. The following lemma shows that this inner ball already belongs to $\BB_0$. 

\begin{lemma}\label{lem:principal-shrinking}
These children satisfy the shrinking and doughnut-leaf properties. Moreover, $\bIN$ belongs to $\BB_0$.
\end{lemma}

\begin{proof}
By Eq.~\eqref{eq:principal-isolated-cluster},
\[
    S\cap\gamma b ~ \subseteq ~ B(x,a^+_u) ~ = ~ \frac{1}{\beta} b_I.
\]
The balls are concentric and $x \in S$, so the doughnut separation condition holds. Furthermore,
\[
    \rad(b_I) ~ = ~ \beta a^+_u ~ \leq ~ \frac{r}{16} 
    \qquad\text{and}\qquad
    4\beta a^+_u ~ \leq ~ \rIN ~ < ~ 8\beta a^+_u ~ \leq ~ \frac r2. 
\]
Thus $b_I \subseteq \bIN$, and the two child cells cover $b$ with the required radius bounds.

It remains to verify that $(u,\rIN)$ is canonical and classified as splitting. Since $\beta \geq 2$, we have $8 a^+_u \leq \rIN < r$. Canonicity of $(u,r)$ and Eq.~\eqref{eq:principal-canonical-range}, with only the lower bound needed for the root, imply that $(u,\rIN)$ is canonical. Finally, $\rIN < 8\beta a^+_u$ implies $a^+_u > \rIN/(16\beta)$, so the ball fails the shrinking test and belongs to $\BB_0$. 
\end{proof}

\subsection{Splitting Nodes}\label{sec:splitting-nodes}

Let $b = B(x,r)$ be a ball in $\BB_0$. We first select principal children covering the sites near $b$, then cover the remaining portions of $b$ by empty derived balls.

\subsubsection{Principal Children}\label{sec:principal}

Apply Lemma~\ref{lem:metric-range-sketch} to $S$ with query center $x$, radius $3 r/2$, resolution $r/32$, and supplied site $x$. Let $C\subseteq S$ be the returned sketch. Its sites are separated by at least $\csk r/32$ and lie in $B(x,3 \Csk r/2)$, so Lemma~\ref{lem:doubling-packing} gives $|C| = 2^{O(\DIM)}$.

For each $y\in C$, select the canonical ball
\[
    b\bigl( u(y,r/4), r/4 \bigr).
\]
Let $\BB'$ be the collection of distinct selected balls. Make their nodes principal children of $b$, creating a node if it does not already exist. A selected ball outside $\BB_0$ satisfies the shrinking test, so it is processed as in the preceding subsection.

\begin{lemma}\label{lem:principal-children-cover-nearby-points}
The collection $\BB'$ has size $2^{O(\DIM)}$, all its balls have radius $r/4$, and every site in $S \cap (3/2)b$ lies within distance $r/16$ of the center of a ball in $\BB'$.
\end{lemma}

\begin{proof}
Only the covering assertion requires proof. For $z \in S \cap(3/2)b$, the sketch contains a site $y$ with
$d(z,y) \leq r/32$. Let $v = u(y,r/4)$. By Eq.~\eqref{eq:principal-canonicalization},
\[
    d(z, \rep(v))
        ~ \leq ~ d(z, y) + d(y, \rep(v))
        ~ \leq ~ \frac{r}{32} + \frac{r}{32}
        ~ = ~ \frac{r}{16}.
\]
\end{proof}

\subsubsection{Empty Derived Children}\label{sec:derived-children}

Set $s = r/32$. By Lemma~\ref{lem:hierarch-ball-cover}, construct an $s$-net $\NN$ for $b$ with $|\NN| = 2^{O(\DIM)}$ and $\NN\subseteq B(x,r+s)$. Retain the centers
\[
    \NN' ~ := ~ \left\{y \in \NN \ST d(y,z) \geq r/4 - s, \text{~for every center $z$ of a ball in $\BB'$}\right\}.
\]
For each $y \in \NN'$, create a derived child with ball $B(y,s)$, and let $\BB''$ denote this collection.

\begin{lemma}\label{lem:derived-children-cover-and-emptiness}
The balls in $\BB' \cup \BB''$ cover $b$, and $S \cap 4 b' = \emptyset$ for every $b' \in \BB''$. There are $2^{O(\DIM)}$ children, each of radius in $[r/32,r/4]$. 
\end{lemma}

\begin{proof}
Let $z \in b$ be uncovered by $\BB'$, and choose $y \in \NN$ with $d(z,y) \leq s$. Since all principal children have radius $r/4$, for each of their centers $z'$ we have
\[
    d(y,z') 
        ~ \geq ~ d(z,z') - d(z,y) 
        ~ > ~ \frac{r}{4} - s.
\]
Thus $y \in \NN'$, and $B(y,s)$ covers $z$. For emptiness, suppose that $z \in S\cap B(y, 4 s)$ for a retained center $y$. Then
\[
    d(x,z) 
        ~ \leq ~ r + s + 4 s 
        ~ < ~ \frac{3 r}{2}.
\]
Lemma~\ref{lem:principal-children-cover-nearby-points} supplies a principal-child center $z'$ with $d(z,z') \leq r/16$. 
Hence
\[
    \frac{7 r}{32}
        ~ = ~ \frac{r}{4} - s
        ~ \leq ~ d(y,z')
        ~ \leq ~ 4 s + \frac{r}{16}
        ~ = ~ \frac{6 r}{32},
\] 
a contradiction. The cardinality and radius bounds follow from the construction.
\end{proof}

\subsubsection{Refinement of Derived Children} \label{sec:derived-child-refinement}

Let $b_v = B(y,s)$ be an initial derived child of a splitting ball $B(x,r)$, where $r = 32 s$. Although $S \cap 4 b_v = \emptyset$, its $\gamma$-expansion need not be empty. Apply Lemma~\ref{lem:hierarch-ball-cover} to $b_v$ with resolution $s/\gamma$, and use the resulting covering DAG as the refinement of $b_v$, identifying its root with the initial derived child. Note that nodes may have multiple parents.

\begin{lemma}\label{lem:derived-refinement}
The DAG resulting from the above construction has $2^{O(\DIM)}\gamma^{\DIM}$ nodes and edges. Its internal nodes satisfy the splitting property, and its cells are empty. Every leaf ball $b'$ satisfies
\[
    S \cap \gamma b' ~ = ~ \emptyset
        \qquad\text{and}\qquad
    x ~ \in ~ S \cap 72 \gamma b'.
\]
\end{lemma}

\begin{proof}
By Lemma~\ref{lem:hierarch-ball-cover}, the resulting DAG has $2^{O(\DIM)}\gamma^{\DIM}$ nodes and edges. Every internal ball is covered by $2^{O(\DIM)}$ children of half its radius, so the splitting property holds. Every node center lies in $B(y, 3 s)$, and every node ball has radius at most $s$. Thus, every node ball is contained in $B(y, 4 s)$. Since $S \cap B(y, 4 s) = \emptyset$, all these cells are empty.

If $b' = B(z,\rho)$ is a leaf, then
\[
    d(y,z) ~\leq ~ 3 s,
        \qquad\text{and}\qquad
    \frac{s}{2\gamma} ~ < ~ \rho ~ \leq ~ \frac{s}{\gamma}.
\]
Every site has distance greater than $4 s$ from $y$, and hence greater than $s$ from $z$. Since $\gamma\rho \leq s$, it follows that $S \cap \gamma b' = \emptyset$. Finally, $d(x,y) \leq 33 s$, so 
\[
    d(x,z) ~ \leq ~ 36 s ~ < ~ 72\gamma\rho. 
\]
Since $x \in S$, this proves $x \in S \cap 72\gamma b'$. 
\end{proof}

Store $x$ as the \emph{locality witness} at each leaf. This takes constant additional storage per leaf. Next, we establish the local multiplicity bound for empty simple leaves (property~(\ref{prop:leaf-multiplicity})). 

\begin{lemma}\label{lem:empty-leaf-multiplicity}
The construction satisfies region-DAG property~(\ref{prop:leaf-multiplicity}).
\end{lemma}

\begin{proof}
Fix $x \in \MM$, $s > 0$, and $\kappa \geq 1$. We first bound the number of splitting principal nodes that can generate relevant leaves, and then bound the number of leaves generated by each such node. Every empty simple leaf is generated by refining a derived child of a splitting principal ball. If the principal ball has radius $r$, its initial derived children have radius $r/32$. By Lemma~\ref{lem:hierarch-ball-cover}, each resulting leaf has radius $\rho$ satisfying
\[
    \frac{r}{64\gamma} 
        ~ < ~ \rho 
        ~ \leq ~ \frac{r}{32\gamma}. 
\]
Thus, leaves with radii in $[s,2 s)$ are generated by principal balls with radii in $[32\gamma s,\ 128\gamma s)$. Since principal radii are powers of two, this interval contains only $O(1)$ possible principal radii.

Fix one such radius $r$. Let $B(p,r)$ be a splitting principal ball generating a leaf $B(z,\rho)$ with $d(x,z) \leq \kappa\gamma s$. The initial derived child whose refinement contains this leaf has radius $r/32$ and center $y$ satisfying $d(p,y) \leq 33 r/32$. By the center-containment guarantee of Lemma~\ref{lem:hierarch-ball-cover}, $d(y,z) \leq 3 r/32$. Consequently,
\[
    d(p,z) ~\leq~ \frac{9 r}{8}. 
\]
Since $r \geq 32\gamma s$, it follows that 
\[
    d(x,p)
        ~ \leq ~ d(x,z) + d(z,p)
        ~ \leq ~ \left(\frac{\kappa}{32} + \frac{9}{8}\right)r. 
\]

All principal balls of radius $r$ are canonical, so their net-tree nodes belong to the same cut $\VV_{k(r)}$. Let $u$ and $v$ be distinct nodes of this cut. Their subtree site sets are disjoint, so $\rep(v) \notin S_u$, and $u$ is not the root. By the packing property, Eq.~\eqref{eq:nt-pack}, we have
\[
    d(\rep(u),\rep(v))
        ~ > ~ \cpack \tau^{\lambda(\parent(u))}
        ~ \geq ~ \cpack\tau^{k(r)}
        ~ > ~ \frac{\cpack}{8\ccov} \, r, 
\]
where the last inequality follows from Eq.~\eqref{eq:principal-cut-level}. Thus, the relevant principal centers lie in a ball of radius $O(\kappa r)$ and are separated by an absolute constant times $r$. Lemma~\ref{lem:doubling-packing} bounds their number by 
\[
    2^{O(\DIM)} \kappa^{\DIM}. 
\]
If the cut consists only of the root, the same bound is immediate.

Each canonical principal ball is stored as a single node and processed only once, regardless of its number of incoming edges. Each splitting principal node creates $2^{O(\DIM)}$ initial derived children. By Lemma~\ref{lem:derived-refinement}, each child's refinement DAG contains $2^{O(\DIM)} \gamma^{\DIM}$ nodes. Hence, each principal node generates at most $2^{O(\DIM)} \gamma^{\DIM}$ empty simple leaves, even when we count separately the leaves in each initial derived child's refinement. Summing over the relevant principal nodes and the $O(1)$ possible principal radii gives at most
\[
    2^{O(\DIM)} (\kappa\gamma)^{\DIM} 
\]
leaves of the required sizes and locations. This proves the property.
\end{proof}

\subsection{The Root Ball} \label{sec:root-ball}

Let $u$ be the root of $T$, and choose
\[
    r_0 ~ := ~ 2^{\ceil{\log_2 \bigl( \max\{8, \beta\} a^+_u \bigr)}} 
    \qquad\text{and}\qquad
    b_0 ~:=~ B(\rep(u),r_0).
\]
Since $n \geq 2$, the root is internal and $a^+_u>0$. The inequality $r_0 \geq 8 a^+_u$ implies $\lambda(u) < k(r_0)$, so $(u,r_0)$ is canonical. Moreover,
\[
    r_0 
        ~ < ~ 2 \max\{8, \beta\} a^+_u 
        ~ < ~ 16 \beta a^+_u, 
\]
and hence $b_0 \in \BB_0$. Finally,
\[
    S 
        ~ \subseteq ~ B(\rep(u), a^+_u) 
        ~ \subseteq ~ \frac{1}{\beta} b_0.
\]
Designate the node corresponding to $b_0$ as the region-DAG root.

\subsection{Scale-Sensitive Point Location}\label{sec:point-location}

A descent from the root may encounter too many scales. Instead, we use a factor-$2$ approximate nearest neighbor and the requested scale $\Lambda$ to choose a query scale and locate a stored splitting ball. If this ball is too large relative to the query scale, one of its principal children answers the query either directly or through a shrinking step. 

\paragraph{Preprocessing.} 
For each net-tree node $u$, the radii of its balls in $\BB_0$ form at most two blocks of consecutive powers of two, since within the canonical range, splitting occurs when $r < 16\beta a^+_u$ or $r > a^-_u/\gamma$. Store each block in an array of node pointers indexed by exponent offset. Arithmetic indexing in the two blocks finds the smallest stored radius at least a given dyadic radius in constant time. These arrays use $O(n\log(\beta\gamma))$ storage in total. 
Also preprocess $S$ for factor-$2$ approximate nearest-neighbor queries using Lemma~\ref{lem:net-tree-ann}.

\paragraph{Entry procedure.} 
Given $q \in b_0$ and $0 \leq \Lambda \leq r_0$, where $r_0 = \rad(b_0)$, find a factor-$2$ approximate nearest neighbor $x \in S$ and put $D = d(q,x)$. If $D$ and $\Lambda$ are not both zero, set 
\[
    r ~ := ~ 2^{\ceil{\log_2 (2 \max\{D,\Lambda\})}}.
\]
If $r \geq r_0$, use the root as the entry node. Otherwise, find $u = u(x,r)$ by a scale-location query and let $b(u, R)$ be the smallest-radius ball associated with $u$ in $\BB_0$ whose radius is at least $r$. If $R \leq 16 r$, use this ball as the entry node. 

If $R > 16 r$, select the principal child $b(u,R/4)$. As shown below, this child exists, contains $q$, and is classified either as singleton or shrinking. If it is singleton, return it as the answer. Otherwise, let $b_I$ be the inner ball of its doughnut child. If $q \notin b_I$, return that doughnut leaf. If $q \in b_I$, return the shrinking node and its simple inner child. 

If $D = \Lambda = 0$, set $r=0$ and let $u$ be the net-tree leaf corresponding to $x$. Let $b(u, R)$ be its smallest-radius ball in $\BB_0$, and return its singleton principal child $b(u, R/4)$. 

\paragraph{Descent.} 
If the entry procedure has not already answered the query, start at its entry node and maintain a current node whose cell contains $q$ and whose size is at least $\Lambda$. At a leaf, return the node. At a splitting node, choose any child whose ball contains $q$. If the child's radius is below $\Lambda$, return the current node together with the chosen child. Otherwise descend to the child. At a shrinking node, return the doughnut leaf if $q$ is outside its closed inner ball. Otherwise use the simple inner child, again returning the current node together with this child if the child's radius is below $\Lambda$, and descending otherwise.

\paragraph{Correctness and query time.}
When $r > 0$, the ANN guarantee and the choice of $r$ give 
\begin{equation} 
    D ~ \leq \frac{r}{2} 
        \qquad\text{and}\qquad
    2 \max\{d(q,S), \Lambda\} 
        ~ \leq ~ r 
        ~ < ~ 8 \max\{d(q,S),\Lambda\}. \label{eq:location-query-scale} 
\end{equation} 
We first show that the entry procedure either answers the query or returns a node of size $O(r)$. Locality then bounds the sizes encountered during descent below by $\Omega(r/\gamma)$, leaving only $O(\log\gamma)$ steps. 

\begin{lemma}[Entry for point location]\label{lem:location-entry}
The entry procedure takes $2^{O(\DIM)} \log n$ time. If $r = 0$, it returns a singleton leaf containing $q$. If $r > 0$, it either answers the scale-sensitive point-location query or returns an entry node whose cell contains $q$ and whose size is between $\Lambda$ and $16 r$.
\end{lemma}

\begin{proof}
Suppose first that $r > 0$. If $r \geq r_0$, the root is a valid entry node: its ball contains $q$ and has radius between $\Lambda$ and $r$. 

Otherwise, let $u = u(x,r)$ and write $y = \rep(u)$. Canonicity and Eq.~\eqref{eq:location-query-scale} give 
\begin{equation}
    d(q,y) 
        ~ \leq ~ D + a^+_u
        ~ \leq ~ \frac{r}{2} + \frac{r}{8}
        ~ = ~ \frac{5r}{8}. \label{eq:location-canonical-contains}
\end{equation}
If $b(u,r)$ is splitting, the successor search selects this ball, which is a valid entry node. 

Suppose instead that $b(u,r)$ passes the shrinking test. Then 
\[ 
    16\beta a^+_u\leq r
        \qquad\text{and}\qquad
    \gamma r\leq a^-_u. 
\] 
The node $u$ is not the net-tree root, since otherwise $r < r_0 < 16 \beta a^+_u$, a contradiction. 

Let $R$ be the smallest dyadic radius greater than $a^-_u/\gamma$. We assert that $b(u, R)$ is the ball selected by the successor search. First, $R > r \geq 8 a^+_u$. Choose a site $z \in S_{\parent(u)}\setminus S_u$, which exists because $\parent(u)$ has at least two children. The packing and covering properties give 
\[
    a^-_u 
        ~ < ~ d(\rep(u), z) 
        ~ \leq ~ 2 a^+_{\parent(u)}. 
\]
Since $\gamma \geq 2$, 
\begin{equation} 
    R 
        ~ \leq ~ 2 a^-_u/\gamma
        ~ \leq ~ a^-_u 
        ~ < ~ 2 a^+_{\parent(u)}. \label{eq:location-successor-radius} 
\end{equation} 
Thus Eq.~\eqref{eq:principal-canonical-range} implies that $b(u, R)$ is canonical. Since $\gamma R > a^-_u$, it fails the shrinking test and is therefore splitting. On the other hand, every canonical dyadic radius $t$ with $r \leq t < R$ satisfies 
\[ 
    16 \beta a^+_u ~ \leq ~ r ~ \leq ~ t 
        \qquad\text{and}\qquad
    \gamma t ~ \leq ~ a^-_u, 
\] 
and therefore passes the shrinking test. Hence the successor search selects $b(u, R)$. 

By Eq.~\eqref{eq:location-canonical-contains}, this ball contains $q$ and has radius at least $r \geq \Lambda$. 
If $R\leq16r$, it is a valid entry node. Suppose that $R > 16r$ henceforth. 

We claim that $b(u, R/4)$ is a principal child of $b(u, R)$. Since $u$ is canonical at radii $r$ and $R$, it is also canonical at $R/4$. Apply Lemma~\ref{lem:principal-children-cover-nearby-points} to the site $y$, the center of the parent ball. There is a principal child $b(v,R/4)$ with center $z = \rep(v)$ satisfying 
\[ 
    d(y,z) 
        ~ \leq ~ R/16
        ~ < ~ a^-_u, 
\] 
where the last inequality follows from Eq.~\eqref{eq:location-successor-radius}. The packing property implies $z \in S_u$. Also, $z \in S_v$, and $u$ and $v$ belong to the same net-tree cut at radius $R/4$. Since the cut subtrees partition $S$, we have $v = u$, proving the claim. 

By Eq.~\eqref{eq:location-canonical-contains}, 
\[ 
    d(q,y)
        ~ \leq~ \frac{5r}{8}
        ~ < ~ \frac{R}{4}, 
\] 
so this child contains $q$. Moreover, 
\[
    16\beta a^+_u
        ~ \leq ~ r
        ~ < ~ R/4
        \qquad\text{and}\qquad
    \gamma R/4 ~ \leq ~ a^-_u/2.
\]
Thus $b(u, R/4)$ passes the shrinking test and is singleton or shrinking. Its radius exceeds $4 r \geq \Lambda$, so returning it when singleton is valid. 

If it is shrinking, put $a = a_u^+ > 0$ and $\rIN = 2^{\ceil{\log_2(4\beta a)}}$. Its doughnut leaf has size $R/4 \geq \Lambda$. Hence returning this leaf when $q$ lies in it is valid. Otherwise, $q \in b_I$. Since the center is a site, $d(q,S) \leq \beta a$ and $D \leq 2\beta a$. If $\Lambda \leq \rIN$, then $\max\{D,\Lambda\} \leq \rIN$. As $\rIN$ is dyadic, the definition of $r$ gives $r \leq 2\rIN < 16 \beta a \leq r$, a contradiction. Thus $\rIN < \Lambda$. Returning the shrinking node together with this simple inner child therefore answers the query.

Finally, suppose that $r = 0$. Then $q = x$ and $\Lambda = 0$. Let $u$ be the net-tree leaf corresponding to $x$, 
so $a^+_u = 0$, and let $R$ be the smallest dyadic radius greater than $a^-_u/\gamma$. The same packing and covering argument gives 
\[ 
    R
        ~ \leq ~ a^-_u 
        ~ < ~ 2 a^+_{\parent(u)}. 
\] 
Thus $b(u, R)$ is canonical and is the smallest splitting ball associated with $u$. Since $a^+_u = 0$, the canonical range for $u$ has no positive lower endpoint. Thus $b(u, R/4)$ is also canonical. Applying Lemma~\ref{lem:principal-children-cover-nearby-points} to $y = \rep(u) = q$, as above, shows that $b(u, R/4)$ is an existing principal child. It is classified as singleton because 
\[ 
    a^+_u ~ = ~ 0
        \qquad\text{and}\qquad
    \gamma R/4 ~ \leq ~ a^-_u/2, 
\] 
and it contains $q$ at its center. 

The procedure uses one ANN query, at most one scale-location query, and one constant-time radius lookup. Selecting the required principal child takes $2^{O(\DIM)}$ time by scanning the children of the successor ball. All remaining work is constant, giving the stated bound. 
\end{proof}

Next, we bound the length of the descent from the entry node.

\begin{lemma}[Short descent]\label{lem:location-descent}
If the entry procedure does not answer the query, descent from the returned entry node answers it in $2^{O(\DIM)} \log\gamma$ time. 
\end{lemma}

\begin{proof}
Child covering and the descent rule preserve containment of $q$ and size at least $\Lambda$. Thus every return satisfies the required conditions; when $\Lambda = 0$, only a leaf can be returned. 

By Lemma~\ref{lem:location-entry}, we may assume $r > 0$ and the entry node has size at most $16 r$. Every leaf containing $q$ has size at least 
\[ 
    \frac{d(q,S)}{\cloc \gamma + 1}. 
\] 
For an empty simple leaf, this follows from Eq.~\eqref{eq:empty-leaf-locality} and the triangle inequality. For every other leaf, it follows from its site-centeredness. Child covering and nonincreasing sizes give the same bound for every node containing $q$. Together with the maintained lower bound $\Lambda$ and Eq.~\eqref{eq:location-query-scale}, this implies that every visited node has size greater than 
\[ 
    \frac{r}{8(\cloc \gamma + 1)}. 
\] 

Each descent to a simple child reduces the size by at least a factor of two, and a doughnut leaf terminates the search. The ratio of the upper and lower size bounds is $O(\gamma)$, so descent takes $O(\log\gamma)$ steps. Each step takes $2^{O(\DIM)}$ time.
\end{proof}

Lemmas~\ref{lem:location-entry} and~\ref{lem:location-descent} establish region-DAG property~(\ref{prop:pt-loc}), since their combined query time is at most $2^{O(\DIM)}\log(n\gamma)$.

\subsection{Node Count, Storage, and Construction Time} \label{sec:global-bounds}

We first verify that the generation rule closes after one round of principal-child requests. The family $\BB_0$ contains $O(n \log(\beta \gamma))$ principal balls, all classified as splitting. Each requests $2^{O(\DIM)}$ principal children. Any requested ball outside $\BB_0$ is a singleton leaf or a shrinking node; in the latter case, Lemma~\ref{lem:principal-shrinking} shows that its simple inner child already belongs to $\BB_0$. Thus, no further splitting balls are introduced. Identical canonical requests are identified, and each principal node is processed only once. Hence the principal family is finite and contains $2^{O(\DIM)}n\log(\beta\gamma)$ balls. 

\begin{lemma}\label{lem:region-dag-size}
The construction produces a valid region-DAG structure with $2^{O(\DIM)} n \gamma^{\DIM} \log(\beta\gamma)$ nodes and total storage. The empty-leaf locality constant, $\cloc$, can be taken to be $72$. 
\end{lemma}

\begin{proof}
Each shrinking principal node creates one doughnut leaf. Each splitting principal node creates $2^{O(\DIM)}$ initial derived children. By Lemma~\ref{lem:derived-refinement}, each child's refinement DAG has $2^{O(\DIM)} \gamma^{\DIM}$ nodes and edges. Together with the bound on the principal family, this gives at most 
\[
    2^{O(\DIM)} n \gamma^{\DIM} \log(\beta\gamma)
\]
nodes in total. All nodes have at most $2^{O(\DIM)}$ children, so the edge count satisfies the same asymptotic bound.

The auxiliary net-tree, range-sketch structure, and ANN structure use $2^{O(\DIM)}n$ storage. The scale-location augmentation uses $O(n)$ additional storage, and the splitting-radius arrays use storage $O(n \log (\beta \gamma))$. The principal-node dictionary uses storage linear in the principal-family size. All these quantities fit within the claimed total storage bound.

The leaf and child-covering region-DAG properties follow from the preceding lemmas, and the designated root has the required margin. Along every edge to a simple node, the radius decreases by at least a factor of two, and the remaining edges lead to doughnut leaves. The finite graph is therefore acyclic. Lemma~\ref{lem:derived-refinement} supplies locality with constant $72$ for every empty simple leaf. All other simple leaves are centered at their unique site. Lemma~\ref{lem:empty-leaf-multiplicity} establishes region-DAG property~(\ref{prop:leaf-multiplicity}). Finally, Lemmas~\ref{lem:location-entry} and~\ref{lem:location-descent} establish the scale-sensitive point-location property.
\end{proof}

\begin{lemma}\label{lem:region-dag-construction-time}
The expected construction time for the region-DAG for $n$ sites given separation parameters $\gamma$ and $\beta$ is
\[
    2^{O(\DIM)} n \gamma^{\DIM} \log(\beta\gamma) \log(n\beta\gamma).
\]
\end{lemma}

\begin{proof}
Let $\ell = \log(\beta\gamma)$. Construct the net-tree, range-sketch structure, and ANN structure in expected time $2^{O(\DIM)} n\log n$. Augmenting the net-tree for scale-location queries takes $O(n)$ additional time and storage by Lemma~\ref{lem:scale-location}. Enumerate the $O(n \SP \ell)$ balls of $\BB_0$ using Lemma~\ref{lem:initial-principal-balls}. Maintain a balanced search dictionary of principal nodes keyed by $(u,j)$, where the radius is $2^j$, and process each node once. Distinct keys specify distinct canonical balls, since distinct nodes in the same cut have distinct representatives. The splitting-radius arrays are assembled during enumeration of $\BB_0$, using $O(n \SP \ell)$ additional time and storage. 

For a splitting principal ball $B(x,r)$, the range sketch takes $O(\log n) + 2^{O(\DIM)}$ time, supplying $x\in S$ as the nearby site. There are $2^{O(\DIM)}$ returned sites, each mapped to its canonical cut node in $O(\log n)$ time. Since there are $2^{O(\DIM)} n \SP \ell$ principal nodes, each dictionary operation costs $O(\log n + \DIM + \log \ell)$. There are $2^{O(\DIM)} n \SP \ell$ such operations in total. Each shrinking principal node requires constant geometric work to define its children, apart from dictionary operations. 

For each splitting principal ball, constructing and filtering its ambient net takes $2^{O(\DIM)}$ expected time. Each of its $2^{O(\DIM)}$ initial derived children is refined using Lemma~\ref{lem:hierarch-ball-cover} with radius-to-resolution ratio $\gamma$. Thus, constructing all its derived refinement DAGs takes expected time
\[
    2^{O(\DIM)} \gamma^{\DIM} \log\gamma.
\]

Combining these costs, and absorbing the factor $\DIM$ from dictionary operations into $2^{O(\DIM)}$, gives total expected time at most 
\[
    2^{O(\DIM)} n \SP \ell \left( \log n + \log \ell + \gamma^{\DIM} \log\gamma \right).
\]
Since $\ell = \log(\beta\gamma) \geq 2$ and $\gamma \geq 2$, this is at most 
\[
    2^{O(\DIM)} n \gamma^{\DIM} \log(\beta\gamma) \log(n\beta\gamma),
\]
as desired.
\end{proof}

Together, the preceding lemmas prove Theorem~\ref{thm:region-dag}.

\subsection{Counting Cell Descriptions over Input Subsets} \label{sec:cell-descriptions}

In this section, we present a counting property of the region-DAG that is useful when the site set is chosen from a fixed finite universe. While we do not use it here, this property is useful in the storage analysis of other AVD applications, where the charging argument of Section~\ref{sec:radius-charging} is not applicable. 

Consider a fixed set $U$ of $n$ sites in some doubling metric space, and consider the region-DAG that results for some subset $S \subseteq U$. Although there are exponentially many such subsets, and hence exponentially many region-DAGs for these subsets, the following lemma shows that the number of distinct cell shapes that can occur in these structures grows only polynomially in $n$.

We make the notion of a cell shape formal through the concept of a cell's \emph{description}. The description of a simple cell consists of its center and radius, and the description of a doughnut cell consists of its common center and its outer and inner radii. Note that different descriptions are counted separately, even if they define the same subset of the metric space $\MM$.

\begin{lemma} \label{lem:region-dag-cell-count}
Let $(\MM,d)$ be a metric space of doubling dimension $\DIM$, let $U \subseteq \MM$ be a set of $n \geq 2$ sites, and fix $\beta,\gamma \geq 2$. Fix a doubling oracle whose output depends only on the input center and radius, and use the hierarchical cover construction of Appendix~\ref{app:hierarchical-ball-covers}. The total number of distinct cell descriptions that can occur in the region-DAG construction, over all subsets $S \subseteq U$ with $|S| \geq 2$ and all outcomes of the randomized preprocessing, is at most
\[
    n^5 \, \gamma^{O(\DIM)} \log(\beta\gamma).
\]
This count includes the descriptions of both internal nodes and leaves.
\end{lemma}

Note that the constants hidden in the exponents and in the multiplicative factor are absolute, and the bound is independent of the spread of $U$.
 
\begin{proof}
We begin by bounding the number of possible net-tree levels and principal balls. Consider an internal node $u$ of a net-tree on any eligible subset $S \subseteq U$. Let $v,w$ be distinct children of $u$, and let $x = \rep(v)$ and $y = \rep(w)$. Since $y \notin S_v$, Eqs.~\eqref{eq:nt-cover} and~\eqref{eq:nt-pack} imply that
\[
    \cpack\tau^{\lambda(u)}
        ~ < ~ d(x,y)
        ~ \leq ~ 2\ccov\tau^{\lambda(u)}.
\]
Thus $\lambda(u)$ belongs to the interval
\[
    \left[ \log_\tau\frac{d(x,y)}{2\ccov},\, \log_\tau\frac{d(x,y)}{\cpack} \right).
\]
Since $\ccov$ and $\cpack$ are absolute constants, each pair of distinct points of $U$ permits only $O(1)$ integer levels. Consequently, a set $\mathcal{K}$ of $O(n^2)$ integers contains every internal level arising over all subsets and all net-tree outcomes.

Let $\mathcal R_0$ be the set of dyadic radii in the intervals
\[
    [8\ccov\tau^k,\,16\beta\ccov\tau^k)
    \qquad\text{or}\qquad
    (\cpack\tau^k/\gamma,\,8\ccov\tau^k), \qquad k\in\mathcal{K}.
\]
Empty intervals contribute no radii. Each $k$ contributes $O(\log(\beta\gamma))$ radii, so
\[
    |\mathcal R_0| ~ = ~ O(n^2\log(\beta\gamma)).
\]
The splitting-band argument in the proof of Lemma~\ref{lem:initial-principal-balls} shows that every initially retained splitting radius belongs to $\mathcal R_0$. The first band uses the level of the node, and the second uses its parent's level. For a net-tree leaf, the first band is absent, and for the root, the second is absent.

Every retained principal ball outside the initial splitting family is requested as a principal child of a splitting principal ball and has one quarter of its radius (Section~\ref{sec:principal}). The simple inner child of a shrinking node already belongs to the initial splitting family (Lemma~\ref{lem:principal-shrinking}). Hence all principal radii belong to
\[
    \mathcal R_0 \cup \{r/4 \ST r\in\mathcal R_0\}.
\]
Every principal center is a site of $U$. It follows that the number of possible principal-ball descriptions is
\begin{equation}
    O(n^3 \log(\beta\gamma)). \label{eq:count-principal-descriptions}
\end{equation}
This argument also covers principal nodes not reachable from the designated root. Range-sketch choices introduce no other centers, since the sketches return input sites.

Next, let us bound the number of doughnut descriptions. The outer ball of a doughnut is a principal ball. Its inner radius is $\beta a_u^+ = \beta \ccov \tau^{\lambda(u)}$ for an internal net-tree node $u$, by Section~\ref{sec:shrinking-nodes}. There are $O(n^2)$ possible inner radii. Multiplying this number by Eq.~\eqref{eq:count-principal-descriptions} bounds all doughnut descriptions by
\begin{equation}
    O(n^5 \log (\beta\gamma)). \label{eq:count-doughnut-descriptions}
\end{equation}

Finally, we bound the number of descriptions for derived balls. We first bound all possible ball descriptions in a hierarchical cover rooted at one fixed parameterized ball $B(z,r)$, with resolution $0 < \rho \leq r$. The construction in Appendix~\ref{app:hierarchical-ball-covers} has $h = \ceil{\log_2(r/\rho)}$ levels of refinement. At a level whose parent radius is $t$, it applies the doubling oracle twice to each parent and obtains at most $Q$ candidate centers, where $2 \leq Q \leq 2^{c\cdot\DIM}$, for an absolute constant $c$. The grouping operation selects centers from these candidates; it generates no new points.

Starting with $B(z,r)$, recursively include every ball $B(q,t/2)$, where $q$ ranges over the centers of the quarter-radius balls obtained by applying the oracle twice to a possible parent $B(x,t)$. Ignore filtering and grouping in this overcount. Since the oracle returns the same output for a given input ball, this is a fixed family with at most $Q$ possible successors per parent. Inductively, every ball occurring at level $j$ of any execution of the actual cover routine belongs to this family at level $j$. Grouping across an entire level does not affect the inclusion, since each selected center is generated by some parent at that level. Assignments may change edges, but do not introduce new centers.

The number of possible descriptions through depth $h$ is therefore at most
\[
    \sum_{j=0}^{h} Q^j 
        ~ \leq ~ Q^{h+1}
        ~ = ~ 2^{O(\DIM)} (r/\rho)^{O(\DIM)}.
\]
The same bound applies to all possible net centers extracted from the terminal level.

For an initially retained splitting ball of radius $r$, the initial derived centers are obtained from a cover at resolution $r/32$ (Section~\ref{sec:derived-children}). Thus there are at most $2^{O(\DIM)}$ possible initial derived balls, each of radius $r/32$. Each such ball is refined at resolution equal to its radius divided by $\gamma$ (Section~\ref{sec:derived-child-refinement}). Since $\gamma \geq 2$, the factor $2^{O(\DIM)}$ is absorbed into $\gamma^{O(\DIM)}$. The preceding bound therefore gives at most $\gamma^{O(\DIM)}$ possible derived-ball descriptions per splitting principal ball, over all preprocessing choices. By Eq.~\eqref{eq:count-principal-descriptions}, the total number of distinct derived-ball descriptions is at most
\[
    n^3 \, \gamma^{O(\DIM)} \log(\beta\gamma).
\]
Every simple node is either principal or derived, and Eq.~\eqref{eq:count-doughnut-descriptions} accounts for all doughnut descriptions. Adding the bounds proves the lemma.
\end{proof}

\section{Approximate Voronoi Diagrams} \label{sec:avd}

Having laid the necessary groundwork for the region-DAG, we are now in a position to construct the AVD and prove Theorem~\ref{thm:avd}. We begin by establishing the tradeoff in part~(\ref{item:avd-tradeoff}) of the theorem, and then we derive the single-representative construction in part~(\ref{item:avd-single-rep}). Throughout, $(\MM,d)$ is a metric space of finite doubling dimension $\DIM$ in the weakly explicit model, $S \subseteq \MM$ is a set of $n \geq 2$ sites, and $0 < \eps \leq 1/2$. The existence arguments use auxiliary ambient-space points permitted by the weakly explicit model; the efficient construction algorithms obtain these points using a doubling oracle. Note that queries will be answered without the use of the oracle.

Recall that an $\eps$-AVD for $S$ consists of a finite searchable cover of $\MM$ by regions, called \emph{cells}, where each cell stores a set of nearest-neighbor representatives (not to be confused with the net-tree's node representatives). Given a region $W \subseteq \MM$, a set $R \subseteq S$ is an \emph{$\eps$-representative set for $W$} if
\[
    d(q,R) ~ \leq ~ (1 + \eps)d(q,S), \qquad\text{for every $q \in W$}.
\]
 A query locates a cell containing the query point and returns a nearest site among that cell's representative set. The storage includes the point-location structure and all representative lists. The point-location structure may use auxiliary regions that are not themselves AVD cells.

Our construction begins by building a region-DAG for $S$, with suitably chosen parameters $\gamma$ and $\beta$. Letting $b_0$ denote the root ball of this region-DAG, we call $\MM \setminus b_0$ the \emph{exterior region}. The region-DAG's leaf cells and the exterior region will be called \emph{terminal regions} and will serve as the cells of our AVD. They cover $\MM$ by region-DAG property~(\ref{prop:covering}). It remains to assign an $\eps$-representative set to each terminal region and bound the resulting storage, query time, and construction time. 

\subsection{Singleton Representatives} \label{sec:radius-singletons}

We first handle all terminal regions except empty simple leaves.

\begin{lemma}\label{lem:radius-singletons}
Suppose $\gamma \geq 2$ and $\beta \geq 1 + 1/\eps$. For every nonempty simple leaf, every doughnut leaf, and the exterior region, the singleton consisting of its defining center is an $\eps$-representative set for that region. For a nonempty simple leaf, the center is an exact nearest neighbor throughout the leaf.
\end{lemma}

\begin{proof}
Consider a nonempty simple leaf $b=B(z,r)$. Property~(\ref{prop:sep-leaf}.\ref{prop:sep-simple}) gives $S \cap \gamma b = \{z\}$. For every $q \in b$ and every $x \in S \setminus \{z\}$,
\[
    d(q,x)
        ~ \geq ~ d(z,x)-d(q,z)
        ~ > ~ (\gamma-1)r
        ~ \geq ~ r
        ~ \geq ~ d(q,z)
\]
(see Figure~\ref{f:conc-ball}(a)). Thus, $z$ is an exact nearest neighbor throughout $b$.

Next, consider a doughnut leaf
\[
    b_O \setminus b_I
        ~ = ~ B(z,r_O) \setminus B(z,r_I)
\]
(see Figure~\ref{f:conc-ball}(b)). By property~(\ref{prop:sep-leaf}.\ref{prop:sep-doughnut}), $z \in S$ and $S \cap \gamma b_O \subseteq \beta^{-1} b_I$. Fix $q$ in this region and put $t = d(q,z)$, so that $r_I < t \leq r_O$. Every site $x \in S$ either belongs to $\beta^{-1} b_I$ or lies outside $\gamma b_O$. In the first case,
\[
    d(q,x)
        ~ \geq ~ t-\frac{r_I}{\beta}
        ~ \geq ~ \left(1-\frac{1}{\beta}\right)t.
\]
In the second case,
\[
    d(q,x)
        ~ > ~ \gamma r_O - t
        ~ \geq ~ (\gamma - 1)t
        ~ \geq ~ t.
\]
Hence,
\[
    d(q,z)
        ~ \leq ~ \frac{\beta}{\beta - 1}d(q,S)
        ~ \leq ~ (1 + \eps) d(q,S).
\]

\begin{figure}[htbp]
    \centering\includegraphics[scale=0.40]{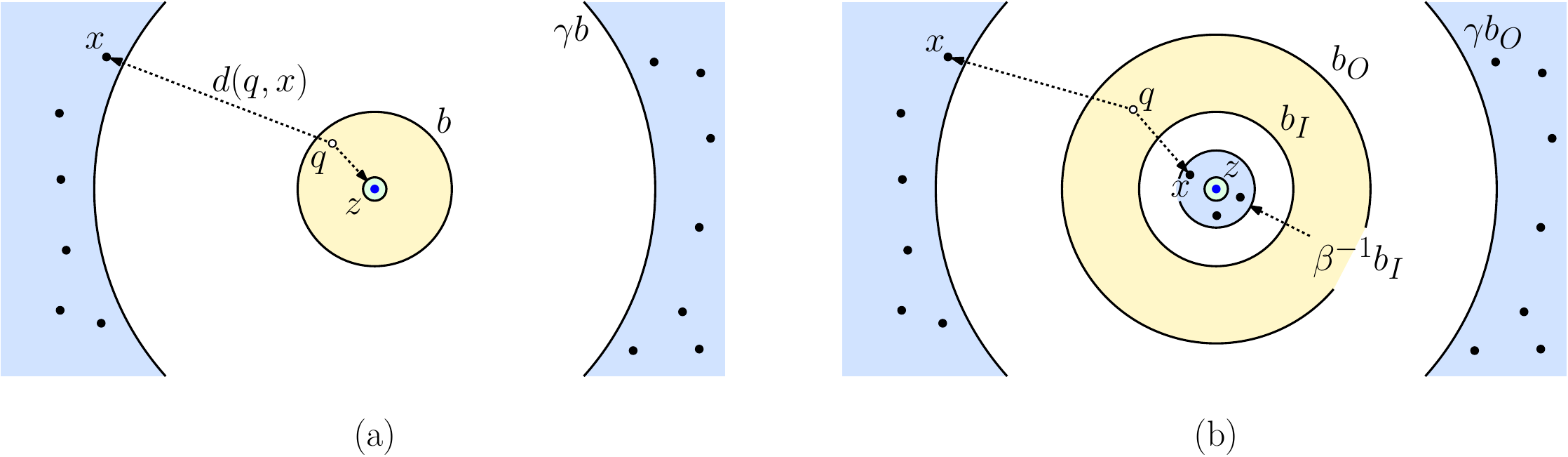}
    \caption{Singleton representatives for terminal regions.} \label{f:conc-ball}
\end{figure}

Finally, let $b_0 = B(z, r_0)$ be the root ball. By property~(\ref{prop:root-ball}), $z \in S$ and $S \subseteq \beta^{-1} b_0$. For $q$ in the exterior region, put $t = d(q,z) > r_0$. For every site $x \in S$,
\[
    d(q,x)
        ~ \geq ~ t - \frac{r_0}{\beta}
        ~ \geq ~ \left(1 - \frac{1}{\beta}\right)t.
\]
The same calculation proves the assertion for the exterior.
\end{proof}

\subsection{Simple Leaves: Space--Time Tradeoff} \label{sec:radius-tradeoff}

Given the results of the previous section, it suffices to focus on representative sets for empty simple leaves. Before considering the single-representative case, we first present the space--time tradeoffs of Theorem~\ref{thm:avd}(\ref{item:avd-tradeoff}).

Fix $2 \leq \gamma \leq 1/\eps$, and set $\beta := 1 + 1/\eps$. Construct the region-DAG on $S$ with outer-separation parameter $\gamma$ and inner-separation parameter $\beta$. Then $\log(\beta\gamma) = O(\log(1/\eps))$. Writing $\LL$ for the set of region-DAG leaves, Theorem~\ref{thm:region-dag} gives
\begin{equation}
    |\LL| ~ \leq ~ 2^{O(\DIM)} n \gamma^{\DIM} \log(1/\eps). \label{eq:radius-dag-size}
\end{equation}
Store the defining center as the sole representative of each nonempty simple leaf, each doughnut leaf, and the exterior region. Let $\EE$ be the set of empty simple leaves, and write $b_w = B(z_w,r_w)$ for the ball associated with each $w \in \EE$.

The following bounds the distances from an empty simple leaf to the sites and all possible nearest neighbors of points in the leaf. 

\begin{lemma}\label{lem:radius-leaf-geometry}
For every $w \in \EE$ and every $q \in b_w$,
\begin{equation}
    \gamma r_w ~ < ~ d(z_w,S) ~ \leq ~ \cloc \gamma r_w
        \qquad\text{and}\qquad
    d(q,S) ~ \geq ~ \frac{\gamma r_w}{2}. \label{eq:radius-query-distance}
\end{equation}
Every nearest site of every point in $b_w$ lies in $B(z_w, \cnear \gamma r_w)$, where $\cnear = \cloc + 1$.
\end{lemma}

\begin{proof}
The first pair of inequalities follows from the simple-leaf separation property and Eq.~\eqref{eq:empty-leaf-locality}. For every $q \in b_w$,
\[
    d(q,S)
        ~ \geq ~ d(z_w,S) - d(z_w,q)
        ~ > ~ (\gamma-1) r_w
        ~ \geq ~ \frac{\gamma r_w}{2}, 
\]
since $\gamma \geq 2$. If $p$ is any nearest site of $q$, then
\[
    d(z_w,p)
        ~ \leq ~ d(z_w,q) + d(q,S)
        ~ \leq ~ 2 r_w + d(z_w,S)
        ~ \leq ~ (\cloc + 1) \gamma r_w.
\]
\end{proof}

\subsubsection{Candidates and Helpers} \label{sec:radius-candidates}

To compute an empty leaf's nearest-neighbor representatives, we first apply the metric range sketch of Lemma~\ref{lem:metric-range-sketch} to the neighborhood surrounding the leaf's cell to obtain a relatively small subset of sites, called \emph{candidates}. To determine which candidates to add to the representative list, we cover the leaf with a metric net of an appropriate resolution, called the \emph{helper set}. The representative set is formed by taking the closest candidate to each helper.

Consider any empty simple leaf $w \in \EE$, and let $b_w = B(z_w, r_w)$ denote its cell. Set
\[
    r_w^{\mathrm{sk}} ~ := ~ \cnear\gamma r_w
        \qquad\text{and}\qquad
    \eta_w ~ := ~ \frac{\eps\gamma r_w}{6}. 
\]
Clearly, $0 < \eta_w \leq r_w^{\mathrm{sk}}$. By Lemma~\ref{lem:radius-leaf-geometry}, we know that the nearest site to any point in $b_w$ lies within $B(z_w, r_w^{\mathrm{sk}})$.

\begin{figure}[htbp]
    \centering\includegraphics[scale=0.40]{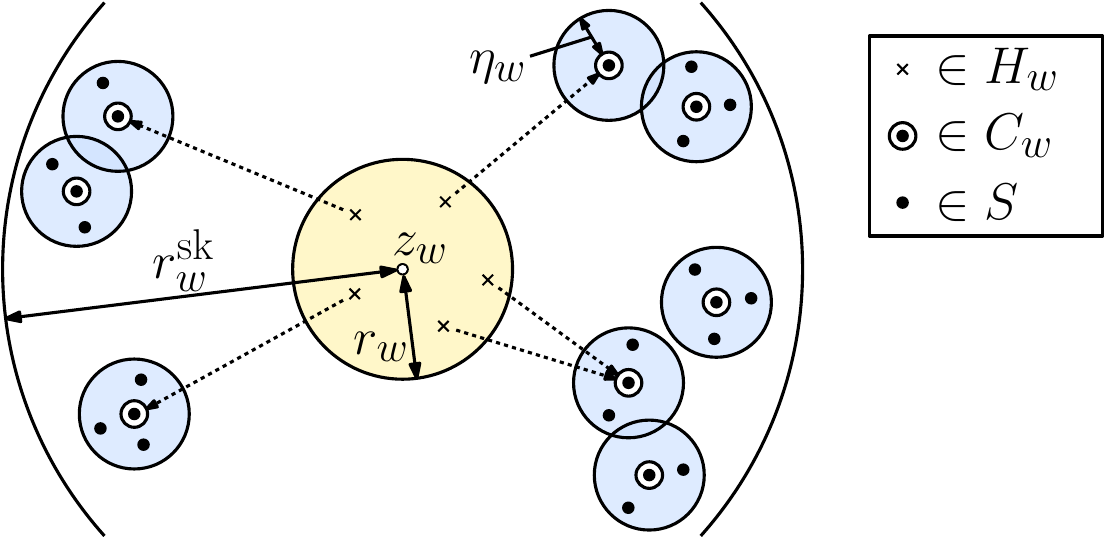}
    \caption{Candidates and helpers.} \label{f:helpers}
\end{figure}

Use the range-sketch structure on $S$ built during the region-DAG construction. Let $p_w$ be the locality witness stored at $w$ (recall Section~\ref{sec:derived-child-refinement}). By property~(\ref{prop:sep-leaf}.\ref{prop:sep-simple}),
\[
    d(z_w, p_w)
        ~ \leq ~ \cloc \gamma r_w
        ~ \leq ~ r_w^{\mathrm{sk}},
\]
so $p_w$ satisfies the nearby-site requirement of the lemma. Query the structure with center $z_w$, radius $r_w^{\mathrm{sk}}$, resolution $\eta_w$, and supplied site $p_w$. The returned candidate set $C_w \subseteq S$ satisfies:
\begin{enumerate}[label=(\roman*)]
\item Every site in $S \cap B(z_w, r_w^{\mathrm{sk}})$ lies within distance $\eta_w$ of a candidate in $C_w$ (see Figure~\ref{f:helpers}).

\item Distinct candidates are separated by at least $\csep \eps \gamma r_w$, where $\csep = \csk/6 > 0$ is an absolute constant. 

\item $C_w \subseteq B(z_w, \Ccand \gamma r_w)$, where $\Ccand = \Csk \cnear$ is an absolute constant. 
\end{enumerate}

By Lemma~\ref{lem:radius-leaf-geometry} and property~(i), every site that is nearest to some point in $b_w$ lies within distance $\eta_w$ of a candidate in $C_w$. In particular, $C_w$ is nonempty.

Apply Lemma~\ref{lem:hierarch-ball-cover} to construct an $\eta_w$-net $H_w$ for $b_w$, whose points we call \emph{helpers}. Thus, every point of $b_w$ lies within distance $\eta_w$ of a helper in $H_w$. Since $\eps \gamma \leq 1$ and $r_w/\eta_w = 6/(\eps\gamma)$, the helper set has size
\begin{equation}
    |H_w| ~ \leq ~ 2^{O(\DIM)} \left(\frac{1}{\eps\gamma}\right)^{\!\DIM}, \label{eq:radius-helper-cardinality}
\end{equation}
and it can be constructed in expected time 
\[ 
    2^{O(\DIM)} \left(\frac{1}{\eps\gamma}\right)^{\!\DIM} \log\left(2+\frac{1}{\eps\gamma}\right). 
\] 

For each $h \in H_w$, select and mark a nearest candidate $x_h$ by scanning $C_w$. A final scan of $C_w$ collects the marked sites into $R_w = \{x_h \ST h\in H_w\}\subseteq C_w$, eliminating repetitions. The total time is $O(|H_w||C_w|+|C_w|)$. Store $R_w$ as the representative list of leaf $w$.

\begin{lemma}\label{lem:radius-representatives}
For every $w \in \EE$, the set $R_w$ is an $\eps$-representative set for $b_w$ and satisfies 
\[ 
    |R_w| 
        ~ \leq ~ |H_w| 
        ~ \leq ~ 2^{O(\DIM)} \left( \frac{1}{\eps\gamma} \right)^{\!\DIM}. 
\] 
\end{lemma}

\begin{proof}
Fix $q \in b_w$ and let $p$ be a nearest site of $q$. By Lemma~\ref{lem:radius-leaf-geometry} and coverage property~(i) of $C_w$, there is an $x \in C_w$ with $d(p,x) \leq \eta_w$. Choose $h \in H_w$ with $d(q,h) \leq \eta_w$. Since $x_h$ is nearest to $h$ among the candidates, 
\[
    d(q, x_h)
        ~ \leq ~ d(q,h) + d(h,x) 
        ~ \leq ~ 2d(q,h) + d(q,p) + d(p,x)
        ~ \leq ~ d(q,p) + 3 \eta_w. 
\]
By Eq.~\eqref{eq:radius-query-distance}, $3\eta_w = \eps \gamma r_w/2 \leq \eps \, d(q,p)$. Therefore,
\[
    d(q,R_w) ~ \leq ~ (1 + \eps) d(q,S).
\]
Finally, $|R_w| \leq |H_w|$, so the cardinality bound follows from Eq.~\eqref{eq:radius-helper-cardinality}.
\end{proof}

\subsubsection{Charging the Candidate Sets} \label{sec:radius-charging}

We next bound the total candidate count, which is used to bound both representative storage and the time spent selecting representatives for the helpers. We adapt the WSPD-based charging argument of~\cite{AMM09a} to the present candidate sets. The argument combines candidate separation and locality with the region-DAG's local-multiplicity property. The first two properties restrict the scales and locations of leaves charging a fixed pair, and the third bounds their number at each scale. 

\begin{figure}[htbp]
  \centering\includegraphics[scale=0.40]{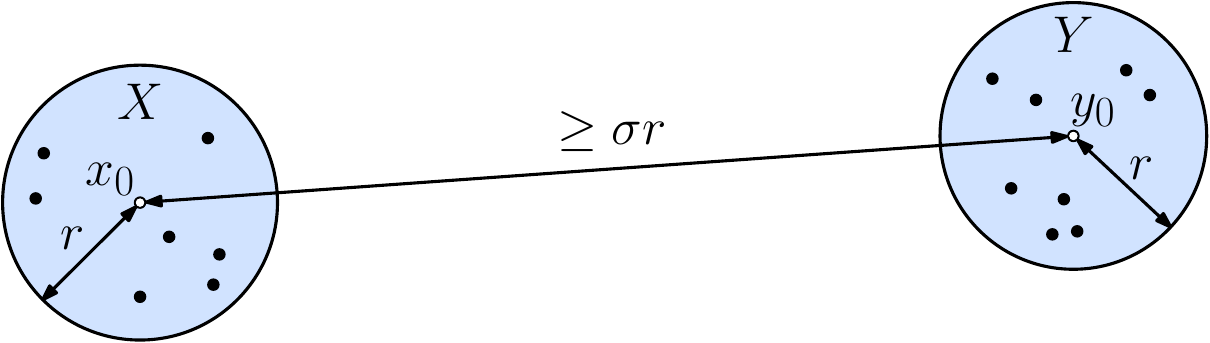}
  \caption{A well-separated pair $(X,Y)$.} \label{f:dumbbell}
\end{figure}

Let us recall some concepts from well-separated pair decompositions (WSPDs). A pair $(X, Y)$ of nonempty disjoint subsets of $S$ is said to \emph{separate} two sites if one belongs to $X$ and the other to $Y$. For $\sigma \geq 2$, the pair is \emph{$\sigma$-well-separated} if there exist equal-radius balls $B(x_0,r)$ and $B(y_0,r)$ containing $X$ and $Y$, respectively, such that
\[
    d(x_0,y_0) ~ \geq ~ \sigma r.
\]
These balls are called the \emph{heads} of the pair (see Figure~\ref{f:dumbbell}). A \emph{well-separated pair decomposition} (WSPD) of $S$ with separation factor $\sigma$ is a collection $\PP$ of $\sigma$-well-separated pairs such that every unordered pair of distinct sites is separated by exactly one pair in $\PP$.

Fix such a decomposition with $\sigma = 16$ and at most $2^{O(\DIM)}n$ pairs. The existence of such a decomposition follows from~\cite[Lemma~5.1]{HaM06}. That construction supplies centers $x_0 \in X$ and $y_0 \in Y$ satisfying
\[
    \max\{\diam(X),\diam(Y)\}
        ~ \leq ~ \frac{d(x_0,y_0)}{16}.
\]
Writing $\ell = d(x_0,y_0)$ for the pair's length, we may therefore take $B(x_0, \ell/16)$ and $B(y_0, \ell/16)$ as the heads. The triangle inequality gives, for every $x \in X$ and $y \in Y$,
\begin{equation}
    d(x,x_0),d(y,y_0) ~ \leq ~ \frac{\ell}{16},
    \qquad\text{and}\qquad
    \frac{7\ell}{8} ~ \leq ~ d(x,y) ~ \leq ~ \frac{9\ell}{8}. \label{eq:radius-wspd}
\end{equation}
The following lemma essentially shows that a WSPD contains a spanning tree over any subset of sites. It uses the closest-pair deletion argument of~\cite[Lemma~7.1]{AMM09a}.

\begin{lemma}\label{lem:radius-witnesses}
For every subset $W \subseteq S$ of cardinality $k \geq 1$, at least $k-1$ distinct WSPD pairs each separate two sites of $W$.
\end{lemma}

\begin{proof}
The assertion is immediate for $k = 1$. Otherwise, repeatedly choose a closest pair among the remaining sites of $W$, record the WSPD pair separating them, and delete one of the chosen sites. Stop when one site remains.

Consider an iteration that chooses $x$ and $y$, records $(X,Y)$ with $x \in X$ and $y \in Y$, and deletes $x$.
By Eq.~\eqref{eq:radius-wspd},
\[
    \diam(X)
        ~ \leq ~ \frac{\ell}{8}
        ~ < ~ \frac{7\ell}{8}
        ~ \leq ~ d(x,y).
\]
If $X$ contained another remaining site $z$, then $d(x,z) < d(x,y)$, contradicting the choice of $x$ and $y$. After deleting $x$, no remaining site belongs to $X$, so $(X, Y)$ cannot be recorded again. The argument is symmetric if the deleted site belongs to $Y$. Thus, the $k-1$ recorded pairs are distinct.
\end{proof}

For each $w\in\EE$, apply the procedure in the proof of Lemma~\ref{lem:radius-witnesses} to $C_w$ and call the resulting $|C_w|-1$ pairs its \emph{witnesses}. The leaf $w$ charges each of its witnesses once. 

\begin{lemma}\label{lem:radius-charges-per-pair}
For every WSPD pair $(X,Y)$, at most $2^{O(\DIM)} \gamma^{\DIM} \log(1/\eps)$ empty simple leaves have candidate sets meeting both $X$ and $Y$. In particular, each WSPD pair receives at most this many witness charges.
\end{lemma}

\begin{proof}
Fix a WSPD pair $(X, Y)$ of length $\ell$ and let $x_0$ be its head center corresponding to $X$. Consider an empty simple leaf $w$ with candidates $x \in C_w \cap X$ and $y\in C_w \cap Y$. Candidate separation and locality imply that
\[
    \csep \eps \gamma r_w 
        ~ \leq ~ d(x,y)
        ~ \leq ~ 2 \Ccand \gamma r_w. 
\]
Combining these inequalities with Eq.~\eqref{eq:radius-wspd} yields
\begin{equation}
    \frac{7\ell}{16\Ccand\gamma} 
        ~ \leq ~ r_w
        ~ \leq ~ \frac{9\ell}{8\csep\eps\gamma}. \label{eq:radius-witness-scales}
\end{equation}
The endpoint ratio is $O(1/\eps)$, so this range is covered by $O(\log(1/\eps))$ intervals of the form $[s,2 s)$.

Fix one such interval. For every relevant leaf $w$ with $s\leq r_w<2 s$, choose $x\in C_w\cap X$. By candidate locality and Eq.~\eqref{eq:radius-wspd}, we have
\[
    d(z_w,x_0)
        ~ \leq ~ d(z_w,x) + d(x,x_0)
        ~ \leq ~ \Ccand\gamma r_w+\frac{\ell}{16} 
        ~ \leq ~ \frac{8\Ccand}{7}\gamma r_w 
        ~ < ~ \frac{16\Ccand}{7}\gamma s, 
\]
where the third inequality follows from the lower bound in Eq.~\eqref{eq:radius-witness-scales}. Apply property~(\ref{prop:leaf-multiplicity}) with center $x_0$, scale $s$, and the absolute constant $\kappa = \max\{1,16\Ccand/7\}$. This bounds the number of these leaf nodes by
\[
    2^{O(\DIM)}(\kappa\gamma)^{\DIM} 
        ~ = ~ 2^{O(\DIM)} \gamma^{\DIM}.
\]
Summing over the $O(\log(1/\eps))$ intervals establishes the claim.
\end{proof}

Let $n_{\mathrm{can}} = \sum_{w\in\EE} |C_w|$ denote the total number of candidates over all empty simple leaves.

\begin{lemma}\label{lem:radius-total-candidates}
The total candidate count satisfies
\[
    n_{\mathrm{can}}
        ~ \leq ~ |\EE| + 2^{O(\DIM)} n \gamma^{\DIM} \log(1/\eps)
        ~ \leq ~ 2^{O(\DIM)} n \gamma^{\DIM} \log(1/\eps).
\]
The representative lists of all the empty simple leaves contain at most $n_{\mathrm{can}}$ entries in total, counting entries in different lists separately.
\end{lemma}

\begin{proof}
For each $w\in\EE$, the candidate set $C_w$ is nonempty, and $w$ charges each of its $|C_w|-1$ selected witness pairs once. For $(X,Y)\in\PP$, let $m_{X,Y}$ be the number of leaves charging $(X,Y)$. Summing the charges received by all WSPD pairs and applying Lemma~\ref{lem:radius-charges-per-pair} and $|\PP| \leq 2^{O(\DIM)} n$, we obtain
\[
    n_{\mathrm{can}} - |\EE|
        ~ = ~ \sum_{(X,Y) \in \PP} m_{X,Y}
        ~ \leq ~ |\PP| \, 2^{O(\DIM)} \gamma^{\DIM} \log(1/\eps)
        ~ \leq ~ 2^{O(\DIM)} n \gamma^{\DIM} \log(1/\eps).
\]
Adding $|\EE|$ proves the first bound. The second follows because $\EE \subseteq \LL$ and Eq.~\eqref{eq:radius-dag-size}. Finally, $R_w \subseteq C_w$ for every $w \in \EE$, so
\[
    \sum_{w \in \EE} |R_w| ~ \leq ~ n_{\mathrm{can}}.
\]
\end{proof}

The witness selection and charges are used only in the analysis and need not be computed during preprocessing.

\subsubsection{Queries and Storage} \label{sec:radius-query-storage}

To answer an approximate nearest neighbor query at $q$, we first test whether it lies in the root ball. If not, we return the root center. Otherwise, we invoke the region-DAG point-location procedure with $\Lambda = 0$ to locate a leaf whose cell contains $q$. For a nonempty simple leaf or a doughnut leaf, return its stored center. For an empty simple leaf $w$, scan its representative list $R_w$, and return the closest site to $q$.

Correctness follows from Lemmas~\ref{lem:radius-singletons} and~\ref{lem:radius-representatives}. Each leaf's representative set is valid throughout its cell, so any leaf whose cell contains $q$ yields a valid answer, even when leaf cells overlap.

By property~(\ref{prop:pt-loc}) and Lemma~\ref{lem:radius-representatives}, the query time is 
\[ 
    2^{O(\DIM)} \left( \log(n\gamma) + \left(\frac{1}{\eps\gamma}\right)^{\!\DIM} \right). 
\] 
Queries are deterministic after preprocessing and do not invoke the doubling oracle. 

The region-DAG and its auxiliary search structures use $2^{O(\DIM)} n \gamma^{\DIM} \log(1/\eps)$ storage by Theorem~\ref{thm:region-dag}. Singleton representatives use $O(|\LL|)$ additional storage. By Lemma~\ref{lem:radius-total-candidates}, the lists of the empty simple leaves contain at most $n_{\mathrm{can}}$ entries in total. Thus, the final storage is
\[
    2^{O(\DIM)} n \gamma^{\DIM} \log(1/\eps) + O(|\LL| + n_{\mathrm{can}})
        ~ \leq ~ 2^{O(\DIM)} n \gamma^{\DIM} \log(1/\eps).
\]

\subsubsection{Construction Time} \label{sec:radius-construction-time}

By Lemma~\ref{lem:region-dag-construction-time}, the expected region-DAG construction time is bounded by 
\[
    T_{\mathrm{DAG}}
        ~ \leq ~ 2^{O(\DIM)} n \gamma^{\DIM} \log(\beta\gamma) \log(n\beta\gamma).
\]
Fix the constructed region-DAG and its search structures, and set $t = 1/(\eps\gamma) \geq 1$. By Lemma~\ref{lem:metric-range-sketch}, the range-sketch queries take $O(|\EE| \log n) + 2^{O(\DIM)} |\EE| + O(n_{\mathrm{can}})$ time in total. Constructing the helper sets takes $2^{O(\DIM)} |\EE| t^{\DIM} \log(2+t)$ expected time. Each helper scans $C_w$ once, and a final scan collects the marked candidates. By Eq.~\eqref{eq:radius-helper-cardinality}, the total time for these scans is 
\[ 
    O\left(\sum_{w \in \EE} (|H_w| + 1)|C_w|\right) 
        ~ \leq ~ 2^{O(\DIM)} t^{\DIM} n_{\mathrm{can}}. 
\] 
Since every candidate set is nonempty, $|\EE| \leq n_{\mathrm{can}}$. Including the singleton representatives, the expected time beyond the region-DAG construction is therefore at most 
\begin{equation} 
    O(|\LL|) + 2^{O(\DIM)} n_{\mathrm{can}} 
        \bigl(\log n + t^{\DIM} \log(2+t)\bigr). \label{eq:radius-refined-time} 
\end{equation} 
By Eq.~\eqref{eq:radius-dag-size} and Lemma~\ref{lem:radius-total-candidates}, $|\LL|$ and $n_{\mathrm{can}}$ are both at most $2^{O(\DIM)} n \gamma^{\DIM} \log(1/\eps)$. Moreover, $\beta = \Theta(1/\eps)$ and $2 \leq \gamma \leq 1/\eps$, so 
\[ 
    T_{\mathrm{DAG}} 
        ~ \leq ~ 2^{O(\DIM)} n \gamma^{\DIM} \log(1/\eps) \log(n/\eps). 
\] 
The parameter bounds and the definition of $t$ also give 
\[ 
    \gamma^{\DIM} ~ \leq ~ (1/\eps)^{\DIM}, 
    \qquad 
    \gamma^{\DIM}t^{\DIM} ~ = ~ (1/\eps)^{\DIM} 
    \qquad\text{and}\qquad
    \log(2+t) ~ = ~ O(\log (n/\eps)). 
\] 
Substituting these uniform bounds into Eq.~\eqref{eq:radius-refined-time} and adding $T_{\mathrm{DAG}}$ gives a total expected construction time of at most  
\[ 
    2^{O(\DIM)} n (1/\eps)^{\DIM} \log(1/\eps) \log(n/\eps). 
\] 
All randomness is confined to preprocessing and does not affect query correctness. This proves Theorem~\ref{thm:avd}(\ref{item:avd-tradeoff}). 

At the low-storage endpoint $\gamma = 2$, we can omit the helpers and store $C_w$ itself at each empty simple leaf $w$. By candidate coverage and Eq.~\eqref{eq:radius-query-distance}, every $q \in b_w$ satisfies $d(q,C_w) \leq d(q,S) + \eta_w \leq (1 + \eps)d(q,S)$. Candidate separation and locality, together with Lemma~\ref{lem:doubling-packing}, give $|C_w| \leq 2^{O(\DIM)} (1/\eps)^{\DIM}$, as required for the query bound at this endpoint. Lemma~\ref{lem:radius-total-candidates} gives the same total storage bound as before. Preprocessing now consists only of constructing the region-DAG and reporting the candidate sets, and hence takes $2^{O(\DIM)} n \log(1/\eps) \log(n/\eps)$ expected time. 

\subsection{The Single-Representative Case} \label{sec:single-representative-avd} 

In this section, we present an AVD construction that results in a single representative for each terminal region, thus establishing Theorem~\ref{thm:avd}(\ref{item:avd-single-rep}). The construction follows by applying the general tradeoff with a sufficiently small approximation parameter such that it suffices to retain just one site from each representative list. First, set 
\[ 
    \delta ~ := ~ \frac{\eps}{4 + \eps}
    \qquad\text{and}\qquad
    \gamma ~ := ~ \frac{1}{\delta} ~ = ~ 1 + \frac4\eps. 
\] 
Apply the construction of Section~\ref{sec:radius-tradeoff} with approximation parameter $\delta$ and tradeoff parameter $\gamma = 1/\delta$. For each empty simple leaf $b = B(z,r)$, let $R_b$ be its $\delta$-representative set. Select a site $p_b \in R_b$ nearest to $z$, retain $p_b$, and discard the remaining entries of the list. Since $z \in b$, 
\[ 
    d(z,p_b) ~ \leq ~ (1 + \delta) d(z,S). 
\] 
For every $q \in b$, the triangle inequality gives 
\[ 
    d(q,p_b) 
        ~ \leq ~ r + (1 + \delta) d(z,S) 
        ~ \leq ~ (1 + \delta) d(q,S) + (2 + \delta)r. 
\] 
The separation property $S \cap \gamma b = \emptyset$ implies that $d(q,S) \geq (\gamma - 1) r$. Therefore, 
\[ 
    d(q,p_b) 
        ~ \leq ~ \left( 1 + \delta + \frac{2 + \delta}{\gamma - 1} \right) d(q,S).
\] 
By our choices of $\delta$ and $\gamma$, 
\[ 
    \delta + \frac{2 + \delta}{\gamma-1} 
        ~ = ~ \frac{3\eps}{4} 
        ~ \leq ~ \eps. 
\] 
Thus, the singleton $\{p_b\}$ is an $\eps$-representative set for $b$. For every other terminal region, retain its defining-center representative, which already satisfies the stronger $(1 + \delta)$-approximation guarantee. The resulting terminal regions form an $\eps$-AVD with one representative per cell. 

Since $1/\delta = \Theta(1/\eps)$, the general construction has at most $2^{O(\DIM)} n (1/\eps)^{\DIM} \log(1/\eps)$ cells and uses the same asymptotic storage. Discarding representatives cannot increase storage. Queries use the same point-location procedure and return the sole representative of the located terminal region, taking $2^{O(\DIM)} \log(n/\eps)$ time. 

Applying the uniform preprocessing bound with approximation parameter $\delta = \Theta(\eps)$ gives an expected construction time of at most 
\[ 
    2^{O(\DIM)} n (1/\eps)^{\DIM} \log(1/\eps) \log(n/\eps). 
\] 
Selecting the retained representatives by scanning each list takes time linear in the total number of list entries and does not increase this bound. Therefore, all the conditions of Theorem~\ref{thm:avd}(\ref{item:avd-single-rep}) are satisfied.

\section{Concluding Remarks and Open Problems} \label{sec:conc}

We have established approximate Voronoi diagrams for doubling metric spaces in the weakly explicit model. Our tradeoff shifts the polynomial dependence on $1/\eps$ from query time at the low-storage endpoint to storage at the high-space endpoint. At the latter endpoint, each cell can store a single representative without changing the asymptotic bounds. Both the tradeoff construction and its single-representative variant use the region-DAG's searchable cover by balls and doughnuts, and their point-location structures may store auxiliary ambient-space centers.

Region-DAGs may also be useful when the ambient space admits a representation in an auxiliary space of bounded doubling dimension. Companion work in preparation investigates this approach for visual, geodesic, locally doubling Gromov-hyperbolic spaces. In that setting, lifting sites to a doubling product of the Gromov boundary with $[0,\infty)$ retains radial-scale information lost under boundary projection. The region-DAG is built in this auxiliary space rather than in the original hyperbolic space.

At both endpoints of the tradeoff, the expected construction time with a doubling oracle is within a logarithmic factor of the stated storage bound for fixed doubling dimension. At $\gamma = 2$, this follows by storing the candidate sets directly. At $\gamma = 1/\eps$, Theorem~\ref{thm:avd} gives this bound for both the tradeoff construction and its single-representative variant. Can the same preprocessing guarantee be achieved throughout the tradeoff range?

\section*{AI Statement}

The authors acknowledge the use of generative AI tools in preparing this manuscript, particularly for outlining, generating, and validating technical details, as well as for grammar checking. The authors take full responsibility for its content.


\bibliographystyle{plainurl}
\bibliography{shortcuts, convex}


\appendix

\section{Net-Tree Implementations} \label{app:net-tree-tools}

In this section we provide proofs of results on net-trees beyond those given by Har-Peled and Mendel~\cite{HaM06}, including Lemmas~\ref{lem:scale-location}, \ref{lem:fixed-scale-grouping}, and~\ref{lem:metric-range-sketch}. We use the net-tree notation and properties of Section~\ref{sec:net-trees}. The finite-metric implementations also use relative lists from the augmented construction of Har-Peled and Mendel~\cite[Section~3.4]{HaM06}, described below.

\scaleLocation*

\begin{proof}
Preprocess $T$ in $O(m)$ time and space to store the depth of each node and to support constant-time level-ancestor queries~\cite{BeF04}. These queries identify an ancestor by tree depth rather than by net-tree level.

Given a scale-location query for a site $x$ and level $k$, return the root if its level is below $k$. Otherwise, apply binary search to the tree depths along the path from the root to the leaf corresponding to $x$. Since net-tree levels strictly decrease along this path (and the leaf has level $-\infty$), the search locates the first node $v$ on this path with $\lambda(v) < k$. By definition, $v$ is the required cut node. Because $T$ has $O(m)$ nodes, the path has length $O(m)$; each binary-search step takes $O(1)$ time via a level-ancestor query, so the total query time is $O(\log m)$.
\end{proof}

The fixed-scale grouping construction first consolidates duplicate input locations using the following lemma. 

\begin{lemma}[Duplicate consolidation] \label{lem:duplicate-consolidation} 
Let $(p_i)_{i=1}^m$ be a nonempty indexed family of points in $\MM$, with repetitions allowed. One can construct a set of representative indices $J \subseteq \{1,\dots,m\}$ and a map $\phi \colon \{1,\dots,m\} \to J$ such that: 
\begin{enumerate} \setlength{\itemsep}{-0.5 ex}\setlength{\parsep}{0 pt} 
\item[$(i)$] $p_i = p_{\phi(i)}$ for every $i$, and $\phi(j) = j$ for every $j \in J$, and 
\item[$(ii)$] the locations $p_j$, for $j \in J$, are distinct. 
\end{enumerate} 
The construction also returns, for each $j \in J$, the list of input indices assigned to $j$. The expected construction time is $2^{O(\DIM)} m \log m$. 
\end{lemma} 

\begin{proof}
We eliminate duplicate locations by recursively applying the following extension of the separating-ball procedure of Har-Peled and Mendel~\cite[Lemma~2.4]{HaM06}, where we sample uniformly from the records and count them with multiplicity. In $2^{O(\DIM)} m$ expected time, this procedure constructs a ball with at least $m/2^{1 + O(\DIM)}$ records inside such that either it has zero radius or its factor-$2$ expansion contains at most $m/2$ records. 
If the ball has zero radius, collect all the records at its center, retaining a single copy, and recurse on the remaining items. On the other hand, if the ball has strictly positive radius, partition the records into those inside and outside the expanded ball and recurse on both parts. Observe that with each application of the separating-ball procedure, we either remove a $2^{-O(\DIM)}$ fraction of the records from further recursion or recurse on two families, each containing at most a $1-2^{-O(\DIM)}$ fraction of the records, with no shared locations between them. Therefore, the expected time is $2^{O(\DIM)}m\log m$. 

At each retained location, keep one original input index as its representative and retain the list of all input indices at that location. These lists define the map $\phi$ and have total size $m$. 
\end{proof}

\fixedScaleGrouping*

\begin{proof}
Apply Lemma~\ref{lem:duplicate-consolidation} to eliminate duplicate locations, retaining the list of original input indices at each location. Let $P$ be the resulting set of distinct locations. Build a net-tree on $P$, storing the corresponding input indices at its leaves, and choose an integer $k$ such that
\[
    \ccov\tau^k 
        ~ \leq ~ s
        ~ < ~ \ccov\tau^{k+1}. 
\]
For each node of $\VV_k$, select an input index whose location is its representative, and let $A$ be the set of selected indices. Assign every input occurrence to the selected index of its cut subtree. In particular, every selected index is assigned to itself. Since each cut node has level below $k$, Eq.~\eqref{eq:nt-cover} gives assignment distance at most $\ccov \tau^k \leq s$, which establishes~(i). Distinct cut nodes have disjoint subtree site sets and parent levels at least $k$. Thus, Eq.~\eqref{eq:nt-pack} implies that the separation is at least
\[
    \cpack \tau^k
        ~ > ~ \frac{\cpack}{\tau \ccov} s. 
\]
Thus, by taking $\cgrp := \cpack/(\tau \ccov)$, we establish~(ii).

A traversal identifies the cut and assigns all input occurrences in $O(m)$ time. Together with duplicate consolidation and net-tree construction, this gives the claimed expected running time.
\end{proof}

To support metric range-sketch queries, we first show how to cover the sites in a query ball by a small family of subtrees without visiting every intervening scale. For each internal node $u$, we use a collection of neighboring nodes at a common cut scale, called the \emph{relative list} of $u$ and defined by
\[
    \Rel(u)
        ~ = ~ \left\{ v \ST
            \begin{array}{l}
                \lambda(v) \leq \lambda(u) < \lambda(\parent(v)) ~~\text{and} \\
                d(\rep(u),\rep(v)) \leq \crel \tau^{\lambda(u)}
            \end{array}
            \right\},
\]
where $\crel$ is an absolute constant. It can be shown that these lists have size $2^{O(\DIM)}$ \cite{HaM06}. We use the constants supplied by Har-Peled and Mendel's construction, which satisfy
\[
    \crel ~ > ~ 2 \ccov + 3.
\]
Indeed, for $\tau = 11$, their constants are $\ccov = 11/5$ and $\crel = 13$ \cite[Definition~2.1 and Section~3.4]{HaM06}. Since levels are integers, the nodes in $\Rel(u)$ belong to $\VV_{\lambda(u) + 1}$, rather than $\VV_{\lambda(u)}$.

The relative lists can be computed as part of the net-tree construction described in the preliminaries~\cite[Theorem~3.1 and Section~3.4]{HaM06}. Including these lists, the construction takes $2^{O(\DIM)} m \log m$ expected time, and the tree and lists together occupy $2^{O(\DIM)} m$ storage. The construction uses only distance evaluations, and both bounds are independent of the spread.

\begin{lemma} \label{lem:app-coarse-cover}
Consider a net-tree on a set $P$ of $m$ sites, augmented with relative lists and the scale-location structure. Given $q \in \MM$, $r > 0$, and a site $p \in P$ with $d(q,p) \leq 2 r$, one can find, in $O(\log m) + 2^{O(\DIM)}$ time, a family $\UU$ of at most $2^{O(\DIM)}$ nodes whose rooted subtrees are pairwise disjoint, such that
\begin{equation}
    P \cap B(q,r)
        ~ \subseteq ~ \bigcup_{u \in \UU}P_u
        ~ \subseteq ~ P \cap B(q, \Ccoarse r), \label{eq:app-sandwich}
\end{equation}
where $\Ccoarse \geq 1$ is an absolute constant. For every integer $k$ satisfying $\ccov\tau^k \leq r$, the parent level of every node in $\UU$ is strictly greater than $k$, with the virtual-parent convention for the root (see Section~\ref{sec:net-trees}).
\end{lemma}

\begin{proof}
Choose an absolute constant $\centry > 1$ such that $\cpack\centry ~ > ~ \ccov + 3$. Set $j := \ceil{\log_\tau (\centry r)}$ and 
\[
    \Ccoarse ~ := ~ 2 + (2\ccov + \crel) \centry. 
\]

If the root level is below $j$, take $\UU$ to consist only of the root. Since levels are integers, $\tau^{\lambda(\mathrm{root})} < \centry r$. The root subtree contains $p$ and has diameter at most $2 \ccov \centry r$. Thus, every site lies within distance $\Ccoarse r$ of $q$. The parent-level condition holds by the virtual-parent convention. 

Otherwise, apply Lemma~\ref{lem:scale-location} to obtain the node $v$ with
\[
    p \in P_v
    \quad\text{and}\quad
    \lambda(v) < j \leq \lambda(\parent(v)).
\]
Set $s = \tau^{\lambda(v)}$, with $s = 0$ if $v$ is a leaf. Since levels are integer-valued, $s < \centry r$. We consider two cases, depending on the relationship between $s$ and $r$.

First, if $s < r$, take $\UU = \{v\}$. For every $x \in P \cap B(q,r)$, we have
\[
    d(\rep(v),x)
        ~ \leq ~ d(\rep(v),p) + d(p,q) + d(q,x)
        ~ \leq ~ \ccov s + 3 r
        ~ < ~ (\ccov + 3)r.
\]
The packing radius of $v$ is at least $\cpack\tau^j \geq \cpack\centry r$, which exceeds this bound. Eq.~\eqref{eq:nt-pack} therefore implies $x \in P_v$. Conversely, every $x \in P_v$ satisfies
\[
    d(q,x)
        ~ \leq ~ d(q,p) + d(p,x)
        ~ \leq ~ 2 r + 2\ccov s
        ~ < ~ (2 + 2\ccov) r.
\]
This proves Eq.~\eqref{eq:app-sandwich}. For any $k$ with $\ccov \tau^k \leq r$, we have $\tau^j \geq \centry r > r \geq \tau^k$, since $\ccov \geq 1$. Thus, $\lambda(\parent(v)) \geq j > k$. 

Otherwise, we have $s \geq r$. Take $\UU = \Rel(v)$. These nodes belong to the cut $\VV_{\lambda(v)+1}$, so their rooted subtrees are pairwise disjoint. Given $x \in P \cap B(q,r)$, let $w$ be the node of this cut whose subtree contains $x$. The covering property and $\lambda(w) \leq \lambda(v)$ give
\begin{align*}
    d(\rep(v),\rep(w))
        & ~ \leq ~ d(\rep(v),p) + d(p,q) + d(q,x) + d(x,\rep(w)) \\
        & ~ \leq ~ 2\ccov s + 3 r
          ~ \leq ~ (2\ccov + 3)s \\
        & ~ < ~ \crel s.
\end{align*}
Thus, $w \in \Rel(v)$, proving coverage. For every $w \in \Rel(v)$ and $x \in P_w$,
\begin{align*}
    d(q,x)
        & ~ \leq ~ d(q,p) + d(p,\rep(v)) + d(\rep(v),\rep(w)) + d(\rep(w),x) \\
        & ~ \leq ~ 2 r + (2\ccov + \crel)s \\
        & ~ < ~ \Ccoarse r.
\end{align*}
This proves locality. For any $k$ with $\ccov \tau^k \leq r$, we have $s \geq r \geq \tau^k$, so $\lambda(v) \geq k$. Every node in $\Rel(v)$ has parent level strictly greater than $\lambda(v)$, proving the parent-level condition. 

The scale-location query takes $O(\log m)$ time, and the relative list has size $2^{O(\DIM)}$. This gives the claimed size and running-time bounds. 
\end{proof}

We now use this subtree cover to answer metric range-sketch queries.

\metricRangeSketch*

\begin{proof}
Apply Lemma~\ref{lem:app-coarse-cover} with the supplied site $p_0$ to obtain $\UU$. Choose
\[
    k ~ = ~ \floor{\log_\tau(\eta/\ccov)}, \qquad\text{so that}\qquad
    \ccov\tau^k
        ~ \leq ~ \eta
        ~ < ~ \ccov\tau^{k+1}.
\]
Within each coarse subtree, expand every node whose level is at least $k$, visiting all its children. Stop at nodes whose levels are below $k$ and return their representatives as $C$.

Since $\ccov\tau^k \leq \eta \leq r$, every coarse node has parent level greater than $k$. Each stopping node therefore belongs to the global cut $\VV_k$, whether it is a coarse node or a descendant of one.

The site sets of the stopping nodes partition $\bigcup_{u \in \UU} P_u$. Since each stopping node has level below $k$, Eq.~\eqref{eq:nt-cover} gives a covering radius of at most $\ccov\tau^k \leq \eta$. Together with Eq.~\eqref{eq:app-sandwich}, this proves coverage of $P \cap B(q, r)$. The stopping subtrees are pairwise disjoint, and every stopping node has parent level at least $k$. Hence, Eq.~\eqref{eq:nt-pack} gives separation at least $\cpack \tau^k$ between distinct returned representatives, where
\[
    \cpack \tau^k ~ > ~ \frac{\cpack}{\tau \ccov} \eta.
\]
Thus, take
\[
    \csk ~ = ~ \frac{\cpack}{\tau \ccov}
    \quad\text{and}\quad
    \Csk ~ = ~ \Ccoarse.
\]
Locality follows from Eq.~\eqref{eq:app-sandwich}.

The visited nodes form a forest whose leaves are the stopping nodes. Each expanded node has at least two children in this forest, so the number of expanded nodes is at most the number of stopping nodes. The stopping subtrees are disjoint, so their representatives are distinct and the number of stopping nodes is $|C|$. Thus the traversal takes $O(|C|)$ time, and the coarse-cover query gives a total worst-case time of $O(\log m) + 2^{O(\DIM)} + O(|C|)$. 

The net-tree and relative lists can be constructed in $2^{O(\DIM)} m \log m$ expected time, and the scale-location augmentation takes an additional $O(m)$ time to construct. The resulting data structure occupies $2^{O(\DIM)} m$ storage, including the $O(m)$ storage for the augmentation. 
\end{proof}

\section{Hierarchical Ball Covers} \label{app:hierarchical-ball-covers}

We construct the hierarchical ball cover one level at a time and extract the net from its terminal balls.

\hierarchBallCover*

\begin{proof}
We give the construction using a doubling oracle. For existence in the weakly explicit model, the same argument applies if, instead of calling the oracle, we choose any cover of the given ball by at most $2^{\DIM}$ balls of half its radius. Such a cover exists by the definition of doubling dimension.

Set 
\[
    h ~ := ~ \ceil{\log_2(r/\rho)}
        \qquad\text{and}\qquad
    t_j ~ := ~ r/2^j.
\]
We construct the DAG in levels $0,\dots,h$, with balls of radius $t_j$ at level $j$. Level zero consists only of $b$.

Given the balls at level $j < h$, let $t = t_j$. For each parent $B(x,t)$, apply two rounds of the doubling oracle to
obtain $2^{O(\DIM)}$ balls of radius $t/4$ covering it. Discard every candidate center $p$ with $d(x,p) > 5 t/4$. Its $t/4$-ball is disjoint from the parent, so the remaining candidate balls still cover the parent.
Keep a separate record for each remaining candidate occurrence, together with its parent.

Apply Lemma~\ref{lem:fixed-scale-grouping} to all candidate records at this level with assignment scale $t/4$. 
For every selected center $q$, create one ball $B(q,t/2)$ at level $j + 1$. For each parent, add an edge to every selected ball to which one of its candidate records is assigned, suppressing duplicate edges. (For example, in Figure~\ref{f:hierarch-ball-cover}, the balls of radius $t$ centered at $x_1$, $x_2$, and $x_3$ generate three balls of radius $t/4$ that are closely packed, which are then grouped by Lemma~\ref{lem:fixed-scale-grouping} into a single ball of radius $t/2$, covering all three.)

\begin{figure}[htbp]
  \centering\includegraphics[scale=0.40]{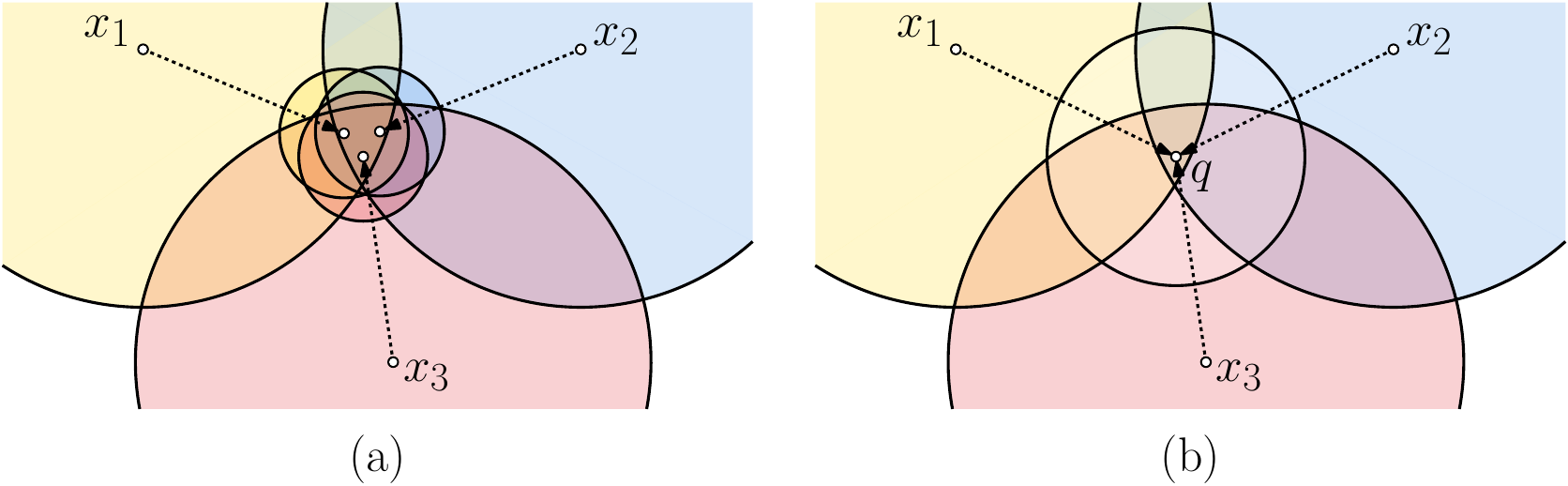}
  \caption{Proof of Lemma~\ref{lem:hierarch-ball-cover}.} \label{f:hierarch-ball-cover}
\end{figure}

Every point of the parent lies within distance $t/4$ of a retained candidate, which is assigned within distance $t/4$ of a selected center. Thus, the children cover the parent. Each parent generates only $2^{O(\DIM)}$ candidate records, so it has at most that many children. For every resulting edge from a parent centered at $x$ to a child centered at $q$,
\[
    d(x,q)
        ~ \leq ~ 5 t/4 + t/4
        ~ = ~ 3 t/2.
\]
Moreover, distinct centers at level $j + 1$ are separated by
at least
\[
    \cgrp t/4
        ~ = ~ (\cgrp/2)t_{j+1}.
\]

Every nonroot ball has an incoming edge, so all nodes are reachable from the root. The distance from $z$ to any level-$j$ center is at most
\[
    \sum_{i=0}^{j-1} \frac{3 t_i}{2}
        ~ = ~ 3 r(1 - 2^{-j})
        ~ < ~ 3 r.
\]
Thus, all node centers lie in $B(z,3 r)$. All leaves are at level $h$ and have the same radius $t_h \in (\rho/2,\rho]$.

Let $n_j$ be the number of balls at level $j$. The separation of their centers and their containment in $B(z,3 r)$ give, by Lemma~\ref{lem:doubling-packing}, 
\[
    n_j 
        ~ \leq ~ 2^{O(\DIM)} (r/t_j)^{\DIM}
        ~ = ~ 2^{O(\DIM)} 2^{j \cdot \DIM}.
\]
Since $\DIM \geq 1$ and $2^h < 2 r/\rho$, summing over the levels bounds the total number of nodes by $2^{O(\DIM)} (r/\rho)^{\DIM}$. Bounded outdegree gives the same asymptotic bound for the number of edges.

We next extract the net without modifying the DAG. The terminal balls cover $b$. Put $\delta = t_h$ and discard every terminal center $q$ with $d(z,q) > r + \delta$. Its $\delta$-ball is disjoint from $b$, so the remaining centers form a set $\NN$ covering $b$ within distance $\delta \leq \rho$. Moreover,
\[
    \NN 
        ~ \subseteq ~ B(z,r + \delta)
        ~ \subseteq ~ B(z,r + \rho).
\]
Distinct points of $\NN$ retain separation of at least
\[
    (\cgrp/2) \delta
        ~ > ~ (\cgrp/4)\rho.
\]
Thus, $\NN$ is a $\rho$-net for the chosen constant $\cnet \leq \cgrp/4$. Its cardinality is at most the number of terminal nodes, and therefore satisfies the claimed bound.

Finally, consider the construction time. At step $j$, the number of candidate records is $m_j \leq 2^{O(\DIM)} n_j$. Oracle calls, filtering, and edge creation take $2^{O(\DIM)} n_j$ time. Grouping takes  expected time $2^{O(\DIM)} m_j \log m_j$. The packing bound gives 
\[
    \log m_j ~ \leq ~ O(\DIM(j + 1)).
\]
Thus, the total expected time to construct the DAG is 
\[
    2^{O(\DIM)} \sum_{j=0}^{h-1} 2^{j \cdot \DIM}(j + 1) 
    ~ \leq ~ 2^{O(\DIM)} (r/\rho)^{\DIM} \log(2 r/\rho).
\]
Here polynomial factors in $\DIM$ are absorbed into $2^{O(\DIM)}$. Extracting the net takes $O(n_h)$ additional time, which is within the same bound. 
\end{proof}

\end{document}